\documentclass[useAMS,usenatbib]{mn2e}

\usepackage{graphicx}
\usepackage{amsmath}
\usepackage{xspace}
\usepackage[utf8]{inputenc}
\usepackage{float}
\usepackage{color}

\usepackage{epstopdf}

\newcommand{\revisions}{}
\newcommand{\revisionstwo}{}

\title[]
{Trans-Dimensional Bayesian Inference for Gravitational Lens Substructures}
    
\author[Brewer, Huijser and Lewis]{%
  Brendon~J.~Brewer$^{1}$\thanks{To whom correspondence should be addressed. Email: {\tt bj.brewer@auckland.ac.nz}},
  David Huijser$^{1}$,
  Geraint F. Lewis$^2$
  \medskip\\
  $^1$Department of Statistics, The University of Auckland, Private Bag 92019, Auckland 1142, New Zealand\\
  $^2$Sydney Institute for Astronomy, School of Physics, A28,
  The University of Sydney, NSW 2006, Australia}

\begin{document}
             
\date{}
             
\maketitle

\label{firstpage}

\begin{abstract}
We introduce a Bayesian solution to the problem of inferring the density
profile of strong gravitational lenses when the lens galaxy may contain
multiple dark or faint substructures. The source and lens models are based on
a superposition of an unknown number of non-negative basis functions
(or ``blobs'') whose form was chosen with speed as a primary criterion.
The prior distribution for the blobs' properties is specified hierarchically,
so the mass function of substructures is a natural output of the method.
We use reversible
jump Markov Chain Monte Carlo (MCMC) within Diffusive Nested Sampling (DNS) to
sample the posterior distribution and evaluate the marginal likelihood of the
model, including the summation over the unknown
number of blobs in the source and the lens.
We demonstrate the method on two
simulated data sets: one with a single
substructure, and one with ten. We also
apply the method to the g-band
image of the ``Cosmic Horseshoe'' system, and find evidence for more
than zero substructures. However, these have large spatial
extent and probably only point to misspecifications in the model
(such as the shape of
the smooth lens component or the point spread function),
which are difficult to guard against in full generality.
\end{abstract}


\section{Introduction}
Galaxy-galaxy gravitational lensing is a powerful astrophysical tool for studying
the distribution of matter, including dark matter, in galaxies
\citep{treu}. One promising application of lensing is to study
and measure the properties of dark matter substructures in the lens
galaxy \citep{koopmans}.
In recent years, this promise has been realised in several lens systems
which are thought to contain at least one dark substructure
\citep{vegetti1, vegetti2, vegetti3}. In these studies, the lens was modelled
as a superposition of a smooth component (such as an elliptical power-law
matter distribution) plus a pixellized ``non-parametric'' correction term.
The estimated spatial structure of the correction term provides clues about
the locations of possible substructures. In a second modelling step, the lens
is assumed to be a smooth overall component plus a compact ``blob'' near any
locations suggested by the correction term. This kind of inference has also
been done with point-like multiple images of {\revisions quasars} \citep{2012MNRAS.419..936F}.

{\revisions In this paper, we introduce} an approach for solving this same problem (inferring the
number and properties of dark substructures given an image) in a more direct
fashion, {\revisions by fitting a model with an unknown number of substructures directly to the image data.}
However, with appropriate modifications, it may also prove useful in other
galaxy-galaxy lens modelling areas. One example is time-delay cosmography, which
requires realistically flexible lens models, and reliable quantification
of the uncertainties
\citep{2013ApJ...766...70S, 2014ApJ...788L..35S, grillo}.

Lens modelling has been a topic of much interest over the last two decades.
Most approaches are based on Bayesian inference \citep{sivia, ohagan}, maximum
likelihood \citep{millar} or
variations thereof. These approaches may be categorized according to the
following criteria:
\begin{enumerate}
\item Whether to use a simply parameterized (e.g. a Sérsic profile),
moderately flexible \citep[e.g. a mixture model,][]{2011MNRAS.412.2521B}
or free-form (e.g. pixellated) model for the surface brightness profile of the source;
\item Whether to use a simply parameterized (e.g. Singular Isothermal Ellipsoid
[SIE]) or flexible model (e.g. pixellated) for the mass profile of the lens; and
\item How to compute the results (e.g. optimization methods, Markov Chain
Monte Carlo [MCMC] with source parameters marginalized out analytically, or
MCMC with the source parameters included).
\end{enumerate}
Recent sophisicated approaches investigating different prior assumptions
and computational approaches include \citet{2014MNRAS.445.2181C},
\citet{2015arXiv150500198T}, and
\citet{2015arXiv150407629B}. 

Each approach involves tradeoffs between convenience and realism.
Simply parameterized models, such as Sérsic surface
brightness profiles for the source,
and singular isothermal ellipsoid plus
external shear (SIE+$\gamma$) models for the lens \citep{1994A&A...284..285K},
are very
convenient. They only have a few adjustable parameters, and they capture
(to ``first order'') relevant prior information that we have about the
source and lens profiles. Of course, they are clearly simplifications,
and can produce misleading results if the actual profile is very different
from any member of the assumed family.

On the
other hand, pixellated models for the source \citep[e.g.][]{suyu} or the lens
\citep[e.g.][]{2014MNRAS.445.2181C} can in principle represent ``any''
source surface brightness profile or lens projected density profile. However, the prior
distribution over pixel values is often chosen to be a multivariate gaussian
for mathematical reasons, so
that the source can be analytically marginalized out
\citep{2003ApJ...590..673W}. Unfortunately, a multivariate gaussian prior
over pixel intensities usually corresponds to a poor model of our prior
beliefs about the source. It assigns virtually zero prior probability
to the hypothesis that the source actually looks like a galaxy, and very high
prior probability to the hypothesis that the source looks like noise (or
blurred noise). These priors also assign nonzero probability to negative
surface brightness or density values; in fact, the marginal prior probability
that any pixel is negative is typically 0.5.

\citet{2011MNRAS.412.2521B} argued that an ideal
modelling approach lies somewhere between simply-parameterised and
pixellated models, which is also a motivation behind
{\revisions the investigation of
shapelets by \citet{2015arXiv150500198T}}.
One way of achieving this is with mixture models.
The source and the lens can be built up from a mixture of a moderate
number (from a few to a few hundred) of
simply parameterised components. For the source, this allows
us to incorporate prior knowledge about the local correlations (the surface
brightness at any particular point is likely to be similar to that at a
nearby point) and the fact that most of the sky is
dark \citep{2006ApJ...637..608B}. The \citet{2011MNRAS.412.2521B} model also
allowed for multi-band data, and allowed for our expectation that the image
may or may not be similar in different bands.

The present paper is similar in approach to \citet{2011MNRAS.412.2521B}.
The main differences are:
i) we use a simpler and faster set of basis functions;
ii) we apply it to the lens as well as the source, allowing for the
possibility of substructure; and
iii) the implementation is based on a C++ template library developed by
\citet{rjobject} which allows for hierarchical priors and trans-dimensional
MCMC to be implemented in a relatively straightforward manner.
However, in the present paper we do not account for
multi-band data; this will be reserved for a future contribution.
The source code for this project is available at
{\tt http://www.github.com/eggplantbren/Lensing2}.

If we are interested in detecting and measuring the properties of
possible dark substructures in a lens galaxy,
a superposition of an unknown number of ``blobs'' is the
most natural model. In addition, if we want to constrain the mass function of
these blobs \citep[e.g.][]{2009MNRAS.400.1583V, 2014MNRAS.442.2017V},
we will need a hierarchical model which specifies the prior probability
distribution for the masses conditional on some hyperparameters. Inferring the
mass function of the blobs then reduces to calculating the posterior
distribution for the hyperparameters. This is related to the general principle
that we should (ideally) construct our inference methods so that the
quantities we infer directly answer our scientific questions
\citep{2015arXiv150507840A, 2015ApJ...807...87S, pancoast}.
{\revisions Here, we are referring to
the mass function of the substructures within a single lens galaxy. An
additional level of hierarchy would be needed to answer questions about a
sample of lens galaxies}.

\section{The Model}
We now describe the details of our model and its parameterization. The
motivation for most of our modelling choices is a compromise between
computational efficiency, realism, and ease of implementation. None of the
choices we have made are final in any sense; rather, this model should be
considered a proof of concept. We encourage exploration of
other choices. Since we can compute marginal likelihoods, Bayesian model
comparison between our choices here and any proposed alternatives should be
straightforward, if a union of our hypothesis space and another
can be considered a reasonable model of prior uncertainty.

\subsection{The Source}
The surface brightness profile of the source is assumed to be composed of
a sum of a finite number of ``blobs'', or basis functions, in order to
allow some flexibility while incorporating prior information about the
non-negativity of surface brightness, and the spatial correlation expected
in real surface brightness profiles.
For computational speed, we choose the following
functional form for the surface brightness profile
of a single blob centered at the origin:
\begin{eqnarray}
f(x, y) &=& \left\{
\begin{array}{lr}
\frac{2A}{\pi w^2}\left(1 - \frac{r^2}{w^2}\right), & r \leq w\\
0, & \textnormal{otherwise}
\end{array}
\right.
\end{eqnarray}
where $r = \sqrt{x^2 + y^2}$, $w$ is the width of the blob, and $A$ is the
total flux of the blob (i.e. the integral of the surface brightness over the
entire domain).
These basis functions are inverted paraboloids, which are faster to evaluate
than gaussians, since they do not
contain an exponential function. In addition, the finite support means that
each blob will evaluate to zero over a large fraction of the domain which
confers an additional speed advantage. {\revisions More conventional choices such as
gaussians are possible, as are alternative compact basis functions, perhaps
inspired by the smoothed particle hydrodynamics literature
\citep{2012MNRAS.425.1068D}.} {\revisionstwo Alternative choices, such as Sérsic
profiles, may be more realistic despite their computational cost. This is
especially likely in the case of an early-type source galaxy
\citep[e.g.][]{eels}. Such modifications are possible without great effort
using the {\tt RJObject} library, and once implemented, the marginal likelihood
can be compared across different choices for the source model.}

If our model contains $N_{\rm src}$ such blobs, positioned at
$\left\{(x_i^{\rm src}, y_i^{\rm src})\right\}$ with widths $\{w_i\}$ and total fluxes
$\{A_i\}$, the overall surface brightness profile is:
\begin{eqnarray}
f(x, y) &=& \sum_{i=1}^{N_{\rm src}}\left\{
\begin{array}{lr}
\frac{2A_i}{\pi w_i^2}\left(1 - \frac{r_i^2}{w_i^2}\right), & r_i \leq w_i\\
0, & \textnormal{otherwise}
\end{array}
\right.
\end{eqnarray}
where $r_i = \sqrt{(x - x_i^{\rm src})^2 + (y - y_i^{\rm src})^2}$.
Under these assumptions, the source can be described in its entirety
by the following parameters:
\begin{eqnarray}
\left\{
N_{\rm src}, \left\{\boldsymbol{\theta}^{\rm src}_i\right\}_{i=1}^N
\right\}
\end{eqnarray}
where $\boldsymbol{\theta}^{\rm src}_i$ denotes the parameters for source blob $i$:
\begin{eqnarray}
\boldsymbol{\theta}^{\rm src}_i &=& \left\{x_i, y_i, A_i, w_i
\right\}.
\end{eqnarray}
The dimensionality of the parameter space for the source depends on the
minimum and maximum values of $N_{\rm src}$, which we set to 0 and 100
respectively (more detail about priors is given in Section~\ref{sec:priors}).
Therefore the source is described by 1 -- 401 parameters.

\subsection{The Lens}
The surface mass density profile of the lens is modelled as a superposition
of a singular isothermal ellipsoid plus external shear (SIE+$\gamma$) and
$N$ circular lens ``blobs''. The SIE+$\gamma$ is a simple and widely used lens
model with analytically available deflection angles, and is intended to account
for the bulk of the lensing effect due to the smooth spatial distribution
of (visible and dark) matter in the lens galaxy. Unfortunately, along with
simplicity comes a lack of realism, and generalising the smooth lens model
is an important future step. {\revisionstwo The softened power law elliptical
mass density or SPEMD model is a popular generalization of the SIE where the
deflection angles can still be computed quickly using the approximations given
by \citet{fastell}.}

Ultimately, concentric
mixtures of smooth components may be the most useful for realistic inference
of the density profile of the halo. A common approach in the lensing
community, that is similar in spirit, is a superposition of an elliptical
power law, external shear, and a pixellated potential correction. However, the
direct use of blobs is closer to the scientific question at hand
when investigating postential dark substructures.

The SIE+$\gamma$ has nine free parameters:
the (circularized) Einstein radius $b$, axis ratio $q$, central position
$(x_c, y_c)$, orientation angle $\theta$, the external shear $\gamma$, and the
orientation angle of the external shear, $\theta_{\gamma}$.

The $N$ lens blobs are intended to model possible dark or faint substructures
in the projected mass profile of the lens.
A lensing blob with mass $M$ and width $v$
centered at the origin has the following surface mass density profile:
\begin{eqnarray}
\rho(x, y) &=& \left\{
\begin{array}{lr}
\frac{2M}{\pi v^2}\left(1 - \frac{r^2}{v^2}\right), & r \leq v\\
0, & \textnormal{otherwise}
\end{array}
\right.
\end{eqnarray}
which is the same as the surface brightness profile of a source blob.
The deflection angles for a single blob are:
\begin{eqnarray}
\alpha_x(x, y) &=&
\left\{
\begin{array}{lr}
\left(2 - r^2/v^2\right)\frac{Mx}{\pi v^2}, & r \leq v\\
\frac{Mx}{\pi r^2}, & \textnormal{otherwise}
\end{array}
\right.\\
\alpha_y(x, y) &=&
\left\{
\begin{array}{lr}
\left(2 - r^2/v^2\right)\frac{My}{\pi v^2}, & r \leq v\\
\frac{My}{\pi r^2}, & \textnormal{otherwise}
\end{array}
\right.
\end{eqnarray}
where $r=\sqrt{x^2 + y^2}$. These do not depend on any
{\revisions computationally expensive} functions such
as square roots or exponentials.
For $N$ such blobs the deflection angles are
summed over all blobs. Therefore the lens can be entirely described by
the following parameters:

\begin{eqnarray}
\left\{
\boldsymbol{\theta}_{\rm SIE},
N_{\rm lens}, \left\{\boldsymbol{\theta}^{\rm lens}_i\right\}_{i=1}^N
\right\}
\end{eqnarray}
where $\boldsymbol{\theta}^{\rm lens}_i$ denotes the parameters for lens blob $i$:
\begin{eqnarray}
\boldsymbol{\theta}^{\rm lens}_i &=& \left\{x_i^{\rm lens}, y_i^{\rm lens}, M_i, v_i
\right\},
\end{eqnarray}
and $\boldsymbol{\theta}_{\rm SIE}$ describes the parameters for the SIE+$\gamma$
component:
\begin{eqnarray}
\boldsymbol{\theta}_{\rm SIE} = \left\{
b, q, x_c^{\rm SIE}, y_c^{\rm SIE}, \theta, \gamma, \theta_\gamma
\right\}.
\end{eqnarray}

Since the model is not of fixed dimension (the number of source and lens
``blobs'' is unknown), we used a trans-dimensional MCMC sampler based on
reversible jump MCMC \citep{green}. This framework has been used successfully
in many astronomical inference problems \citep[e.g.][]{jones, renate}.
We implemented the MCMC using the
{\tt RJObject} software \citep{rjobject}, a C++ library for implementing trans-dimensional MCMC with hierarchically-specified priors when the model is
a ``mixture model'' or similar, as is the case here.
The {\tt RJObject} library uses the
Diffusive Nested Sampling algorithm \citep[DNS;][]{dnest} for its sampling, but the
trans-dimensionality is handled by the MCMC moves. Therefore, the marginal
likelihood we obtain is one that involves a sum over the hypothesis space
for $N_{\rm src}$ and $N_{\rm lens}$. In other words, we do not need
separate runs with different trial values of $N_{\rm src}$ and $N_{\rm lens}$.
Previous astronomical
applications of {\tt RJObject} include \citet{magnetron}
and~\citet{exoplanet}. Proposal moves for the parameters (positions and masses
of the source and lens blobs), and birth/death moves to add or remove blobs
to the source or lens, are handled internally by {\tt RJObject} and
fit into the general framework described in \citet{rjobject}.

\section{Prior Probability Distributions}\label{sec:priors}
\begin{figure*}
\includegraphics{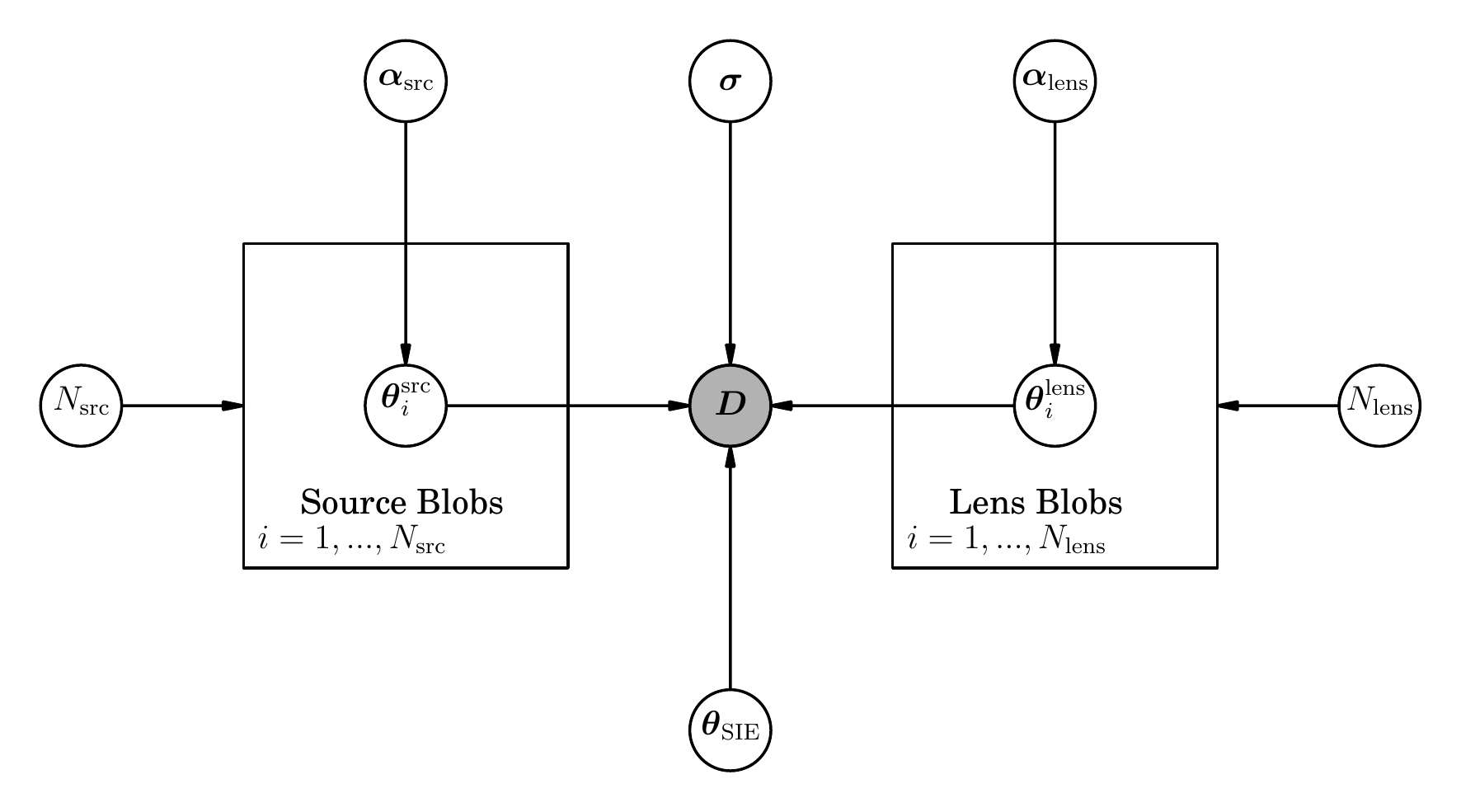}
\caption{A probabilistic graphical model (PGM) of the dependence structure
of the prior information, produced using DAFT ({\tt www.daft-pgm.org}).
The prior for the source and lens blob parameters are specified conditional
on hyperparameters. The purpose of this is to induce dependence in the prior
distribution for the blob parameters, implying (for example) that the mass of
one blob is slightly informative about the mass of another, and that the
locations might be clustered around a certain typical location.
\label{fig:pgm}}
\end{figure*}

The prior probability distributions for all hyperparameters, parameters, and
the data, are given in Table~\ref{tab:priors}, and a factorization of the
joint distribution is displayed in Figure~\ref{fig:pgm}. With these priors,
we aim to express a large degree of prior uncertainty (hence the liberal use
of uniform, log-uniform, and Cauchy distributions). However, we also specify
some priors hierarchically to allow (for example)
blobs to cluster around a typical central
position, rather than implying a high probability for substructure positions
being spread uniformly (in a frequency sense) over the sky. We also applied
hierarchical priors to the substructure masses (and source fluxes) so that, while
the typical order of magnitude of the masses is unknown, the masses
themselves are likely to be roughly the same order of magnitude. The
hyperparameters of these distributions can also be interpreted as
straightforward answers to questions about the substructure mass function.

\begin{table*}
\begin{tabular}{|l|l|l|}
\hline
Quantity	&	Meaning		& Prior\\
\hline
{\bf Numbers of Blobs}\\
\hline
$N_{\rm src}$	&	Number of source blobs	& Uniform($0, 1, ..., 100$)\\
$N_{\rm lens}$	&	Number of lens blobs	& Uniform($0, 1, ..., 10$)\\
\hline
{\bf Source hyperparameters} ($\boldsymbol{\alpha}_{\rm src}$)\\
\hline
$(x_c^{\rm src}, y_c^{\rm src})$ & Typical position of source blobs & iid Cauchy(location=$(0,0)$, scale=0.1$\times${\tt imageWidth})\\
$R_{\rm src}$   &	Typical distance of blobs from $(x_c^{\rm src}, y_c^{\rm src})$ & LogUniform(0.01$\times${\tt imageWidth}, 10$\times${\tt imageWidth})\\
$\mu_{\rm src}$ &	Typical flux of source blobs	& $\ln(\mu_{\rm src}) \sim$ Cauchy(0, 1)$T(-21.205, 21.205)$\\
$W_{\rm max}^{\rm src}$ &		Maximum width of source blobs	& LogUniform(0.001$\times${\tt imageWidth}, {\tt imageWidth})\\
$W_{\rm min}^{\rm src}$ &		Minimum width of source blobs	& Uniform(0, $W_{\rm max}^{\rm src}$)\\
\hline
{\bf Lens hyperparameters} ($\boldsymbol{\alpha}_{\rm lens}$)\\
\hline
$(x_c^{\rm lens}, y_c^{\rm lens})$ & Typical position of lens blobs & iid Cauchy(location=$(0,0)$, scale=0.1$\times${\tt imageWidth})\\
$R_{\rm lens}$  &	Typical distance of blobs from $(x_c^{\rm lens}, y_c^{\rm lens})$ 	& LogUniform(0.01$\times${\tt imageWidth}, 10$\times${\tt imageWidth})\\
$\mu_{\rm lens}$&	Typical mass of lens blobs	& $\ln(\mu_{\rm lens}) \sim$ Cauchy(0, 1)$T(-21.205, 21.205)$\\
$W_{\rm max}^{\rm lens}$ &		Maximum width of lens blobs	& LogUniform(0.001$\times${\tt imageWidth}, {\tt imageWidth})\\
$W_{\rm min}^{\rm lens}$ &		Minimum width of lens blobs	& Uniform(0, $W_{\rm max}^{\rm lens}$)\\
\hline
{\bf Source Blob Parameters} ($\boldsymbol{\theta}_i^{\rm src}$)\\
\hline
$(x_i^{\rm src}, y_i^{\rm src})$ & Blob position & Circular(location=$(x_c^{\rm src}, y_c^{\rm src})$, scale=$R_{\rm src}$)\\
$A_i$  & Blob flux & Exponential(mean=$\mu_{\rm src}$)\\
$w_i$  & Blob width & Uniform($W_{\rm min}^{\rm src}$, $W_{\rm max}^{\rm src}$)\\
\hline
{\bf Lens Blob Parameters} ($\boldsymbol{\theta}_i^{\rm lens}$)\\
\hline
$(x_i^{\rm lens}, y_i^{\rm lens})$ & Blob position & Circular(location=$(x_c^{\rm lens}, y_c^{\rm lens})$, scale=$R_{\rm lens}$) \\
$M_i$  & Blob mass & Exponential(mean=$\mu_{\rm lens}$)\\
$v_i$  & Blob width & Uniform($W_{\rm min}^{\rm lens}$, $W_{\rm max}^{\rm lens}$)\\
\hline
{\bf Smooth Lens Parameters} ($\boldsymbol{\theta}_{\rm SIE}$)\\
\hline
$b$ & SIE Einstein Radius & LogUniform(0.001$\times${\tt imageWidth}, {\tt imageWidth})\\
$q$ & Axis ratio & Uniform(0, 0.95)\\
$(x_c^{\rm SIE}, y_c^{\rm SIE})$ & Central position & iid Cauchy(location=$(0,0)$, scale=0.1$\times${\tt imageWidth})\\
$\theta$ & Orientation angle & Uniform$(0, \pi)$\\
$\gamma$ & External shear & Cauchy$(0, 0.05)T(0, \infty)$\\
$\theta_\gamma$ & External shear angle & Uniform$(0, \pi)$\\
\hline
{\bf Noise Parameters} ($\boldsymbol{\sigma}$)\\
\hline
$\sigma_0$ & Constant component of noise variance & $\ln(\sigma_0) \sim$ Cauchy(0, 1)$T(-21.205, 21.205)$\\
$\sigma_1$ & Coefficient for variance increasing with flux &
$\ln(\sigma_1) \sim$ Cauchy(0, 1)$T(-21.205, 21.205)$\\
\hline
{\bf Data} ($\boldsymbol{D}$)\\
\hline
$D_{ij}$ & Pixel intensities & Normal$(m_{ij}, s_{ij}^2 + \sigma_0^2 + \sigma_1m_{ij})$
\end{tabular}
\caption{The prior distribution for all hyperparameters, parameters, and the
data, in our model. Uniform$(a, b)$ is a uniform
distribution between $a$ and $b$. LogUniform$(a, b)$ is a log-uniform
distribution (with density $f(x) \propto 1/x$, sometimes erroneously called
a Jeffreys prior) between $a$ and $b$. The notation $T(\alpha, \beta)$ after
a distribution denotes truncation to the interval $[\alpha, \beta]$. The
constant {\tt imageWidth} is the geometric mean of the image dimensions in
the $x$ and $y$ directions.
\label{tab:priors}}
\end{table*}

We assign somewhat informative (but heavy-tailed)
Cauchy priors to the central positions of the source and the lens.
One might object to the Cauchy priors
for the central positions (and argue instead for gaussians) on the
basis that they do not have rotational symmetry. However, the priors we
are assigning are prior only to the values of the data pixel intensities only,
and not other facts about the data, such as its dimensions in arc seconds, or its
rectangular shape. Given these, there is no reason to insist on rotational
symmetry. The Cauchy priors are intended to enhance the plausibility
(relative to what a uniform prior would imply) that the
lens and source are somewhere near the centre of the image, but in a cautious
way.

The ``circular'' conditional prior for the blob positions has the following
density:
\begin{eqnarray}
\rho(x, y)\, dx \, dy &=&
\frac{1}{2\pi W} \frac{1}{r}\exp\left(-\frac{r}{W}\right) \, dx \, dy
\end{eqnarray}
where $r = \sqrt{(x-x_c)^2 + (y-y_c)^2}$. This results from an exponential
distribution (with expectation $W$) for the radial coordinate and a uniform distribution for the angular coordinate in plane polar coordinates. This was
chosen (as opposed to a more ``obvious'' choice such as a normal distribution)
for computational reasons: the {\tt RJObject} code requires a function to
transform from Uniform(0, 1) distributions to this distribution and back
(making use of cumulative distribution functions and their inverses), and
the normal would have required the use of special functions for this.

\subsection{Conditional prior for the data}
The conditional prior for the data given the parameters
is a product of independent gaussian distributions, one for each pixel.
The mean of the gaussian is given by the ``mock'' noise-free image we would
expect based on the parameters. The standard deviation of the gaussian is
a combination of three terms; the first (denoted $s_{ij}$) is a ``noise map''
which is loaded from a file, the second is an unknown constant $\sigma_0$
which applies to the whole image, and the third is proportional to the
square root of the mock image, with proportionality constant $\sigma_1$.

This is usually called the ``sampling distribution'', or sometimes just the
``likelihood''. However, sampling distribution is misleading since no
physical frequency distribution exists which is being sampled from. The term
likelihood is also usually used to refer to the scalar function of the parameters
obtained when the data are known. For a discussion of the view that this
object is really a prior distribution, see \citet[][pp. 33--35]{caticha}.

It is common to provide a ``variance image'' to lens modelling software,
which specifies the standard deviations used in the likelihood function.
However, telescopes do not provide variance images. It also does not make
sense to specify vague priors prior to the image pixel values but posterior to
the variance image (which usually resembles the image itself, and therefore
would be highly informative). Allowing the standard deviation of the gaussian in the
likelihood to depend on the mock image brightness is a more principled way
of modelling our actual inferential situation. Nevertheless, we allow for an
input variance image as well, which can be used for ad-hoc masking of
troublesome regions (such as unmodelled flux from other sources).

\section{Computation}
Computing the posterior distribution over the parameters of such a model
requires that we can implement Markov Chain Monte Carlo (MCMC) over the space
of possible sources and lenses.
To compute the posterior distribution for the parameters
(of which there are 24--464, depending on the values of $N_{\rm src}$ and
$N_{\rm lens}$), we use Diffusive Nested Sampling \citep{dnest}, a form
of Nested Sampling \citep{skilling} that uses the Metropolis algorithm
to move around the parameter space.

The proposal distributions for the blob parameters (both source and lens) are
handled by the {\tt RJObject} library \citep{rjobject}. This includes
birth and death proposals that increase or decrease either $N_{\rm src}$,
or $N_{\rm lens}$, as well as proposals that move the blobs (in their parameter
space) while keeping the number of blobs fixed. {\tt RJObject} also
facilitates proposals that change the hyperparameters (either for the lens
or the source) while keeping the actual blobs in place, as well as proposals
that change the hyperparameters and shift all of the relevant blobs in
so they are appropriate for the new values of the hyperparameters.

Evaluating the likelihood function requires that we compute a ``mock'' image from
the current setting of the parameters. This mock image is calculated using
standard ray-tracing methods with a uniform grid of $n \times n$ rays
fired per image pixel. However, certain kinds of proposals do not affect the
image in any way, such as those which change the noise parameters. For efficiency
we do not recompute the mock image in these cases.

\section{{\revisions``Easy''} Simulated Data}
\begin{figure}
\begin{center}
\includegraphics[scale=0.5]{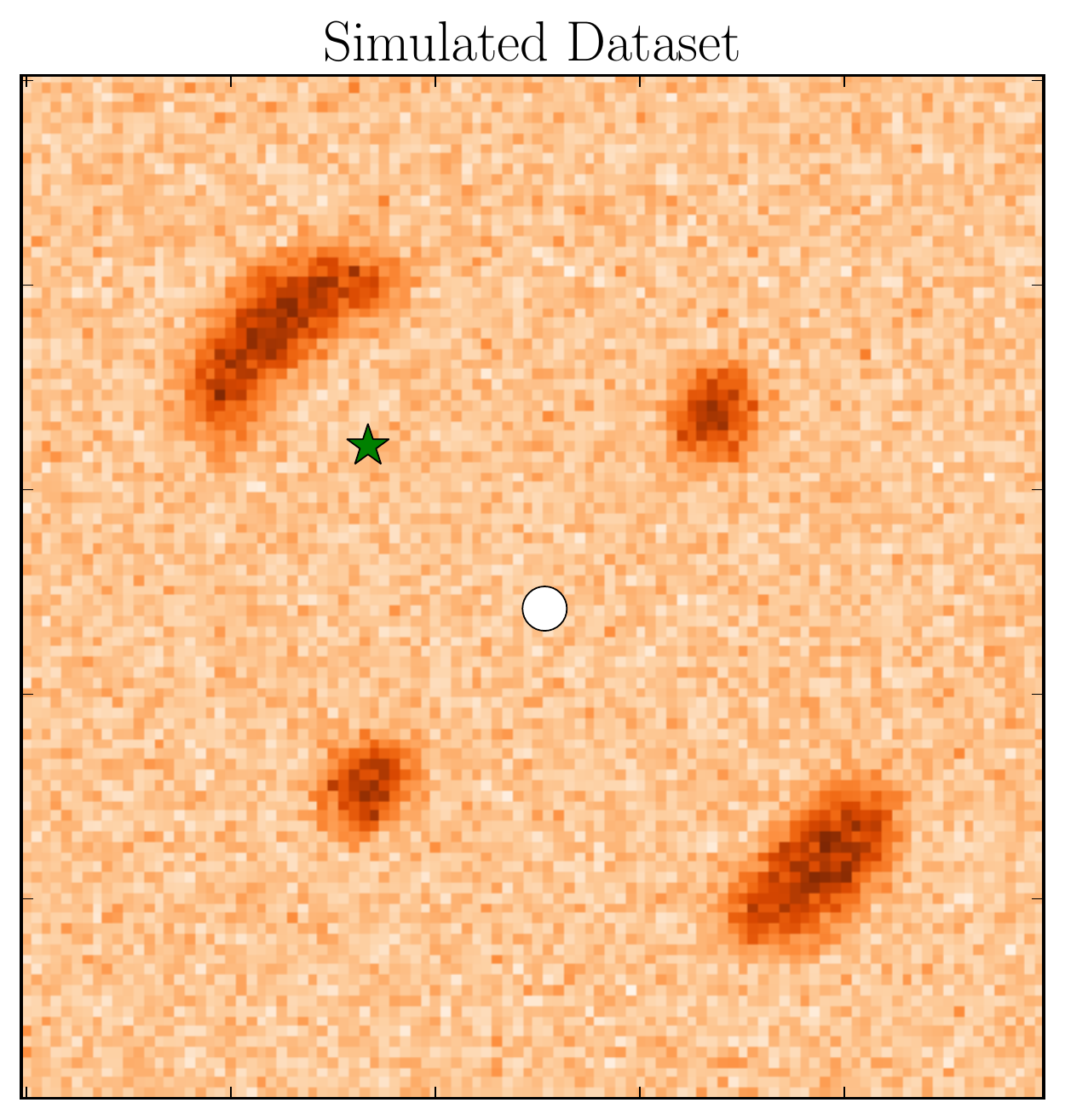}
\caption{A simulated image of a simple ``galaxy'' lensed by an SIE+$\gamma$
lens close to the center of the image, indicated by the white circle, plus a
single substructure close to the top left image (indicated by a star symbol).
The image was blurred by a PSF and had some noise added.
In an arbitrary system of units, this image extends from -8 to 8 in the
$x$ and $y$ directions.\label{fig:image1}}
\end{center}
\end{figure}

To demonstrate the method, we generated a simulated dataset {\revisions where
the lens was an SIE$+\gamma$ profile with a single additional substructure}.
The image is shown in Figure~\ref{fig:image1}, and
consists of 100 $\times$ 100 pixels. The point spread function (PSF) was compact and was
defined on a 5 $\times$ 5 grid. The data was created by firing only one ray
per pixel, and the inference was also carried out under the same level of
approximation. Since the model assumptions are all correct for this image, the
demonstration here is purely to illustrate the computational tractability of
the problem, and the kind of outputs the method can produce.
{\revisions
The SIE$+\gamma$ parameters were, {\revisions to three significant figures},
$\{b, q, x_c^{\rm SIE}, y_c^{\rm SIE}, \theta, \gamma, \theta_\gamma\}
= \{5.25, 0.323, 0.191, -0.341, 0.833, 0.0322, 0.497\}$, where the length
units in the image plane are arbitrary, and rotation angles are measured in
radians. The single substructure is located at $(-2.57, 2.20)$, about halfway
between the center and the upper left image.
Its mass is 2.50 units, using the convention where the
critical density is 1. For comparison, the SIE mass within its critical
ellipse is $\pi b^2 = 86.6$ units.
}

We executed the code to generate 5,000 samples from the ``mixture of
constrained priors'' distribution of DNS. Our thinning factor was $10^5$,
so $5 \times 10^8$ MCMC iterations were actually performed, taking approximately
48 hours on a modest desktop PC\footnote{The computer was purchased in 2012
and has a second-generation intel i7 processor, and the process was run on
8 threads.}. After resampling these samples to
reflect the posterior distribution, we were left with {\revisions 500}
(equally weighted) posterior samples.

The posterior distributions for complex models, such as the mixture models used here, are often challenging and unintuitive to summarise. One effective way
to visualise the uncertainty in the inferences is to play a movie where each
frame is a sample from the posterior distribution. The degree to which the
frames differ from each other conveys the uncertainty remaining after taking
the data into account.

For the purposes of a paper, static summaries are more convenient than movies.
One useful summary is based on the concept of empirical measure.
The {\it empirical measure} of the substructure positions is a function that
takes the actual substructure positions $(x_i^{\rm lens}, y_i^{\rm lens})$ and
produces a ``density function'' over two dimensions, composed of delta
functions at the positions themselves:
\begin{eqnarray}
\left\{
(x_i^{\rm lens}, y_i^{\rm lens})
\right\}_{i=1}^{N_{\rm lens}}
\Longrightarrow
\sum_{i=1}^{N_{\rm lens}}
\delta^2\left(x - x_i^{\rm lens}, y - y_i^{\rm lens}\right).
\end{eqnarray}

Intuitively, the empirical measure is a mathematical object that is like
an ``infinite resolution histogram'', in this case a two dimensional histogram,
of the substructure positions.
Being a function of the actual substructure positions, the empirical measure
is not available to us since we do not know those positions with certainty.
However, we have samples from the posterior distribution for those positions,
and can use these (trivially) to create samples from the posterior distribution
for the empirical measure. We can also summarise this posterior, for example,
by taking its expected value.

The posterior expected value of the empirical measure is:
{\revisions
\begin{eqnarray}
\int p(\boldsymbol{\theta} | \boldsymbol{D})
\sum_{i=1}^{N_{\rm lens}}
\delta^2\left(x - x_i^{\rm lens}, y - y_i^{\rm lens}\right)
\, d\boldsymbol{\theta}
\end{eqnarray}
}
where $\boldsymbol{\theta}$ denotes all parameters and hyperparameters
(including the $N$s) and ``$\int \,d\boldsymbol{\theta}$'' is an integral and summation over the entire parameter space.

Since we can approximate posterior expectations using Monte Carlo, we can
obtain the expected value of the empirical measure using:
{\revisions
\begin{eqnarray}
\frac{1}{n}
\sum_{k=1}^n
\sum_{i=1}^{N_{\rm lens}^k}
\delta^2\left(x - x_{ik}^{\rm lens}, y - y_{ik}^{\rm lens}\right)
\end{eqnarray}
}
where $n$ is the number of posterior samples. The resulting function is
an ``image'' with a point mass wherever a substructure occurred. For
visualisation purposes the image can be blurred, or calculated at a lower
resolution by replacing the Dirac-delta function with a discrete version
which returns a nonzero constant
if a substructure appears in a pixel or zero otherwise.

The masses of the smooth and substructure components of the lens are usually
of interest. Since the total mass of an SIE lens is infinite, the question
needs to be redefined, so we ask about the mass within some aperture of finite
area. For an SIE, the mass (in dimensionless units based on the critical density)
within the critical
ellipse is simply $\pi b^2$. However, the blobs have finite total masses
$\{M_i\}$. To obtain the posterior distribution for the lens masses, one must
be clear about exactly which mass they are talking about, and many definitions
are possible, although some might be more meaningful or well constrained than
others.

\begin{figure}
\begin{center}
\includegraphics[scale=0.4]{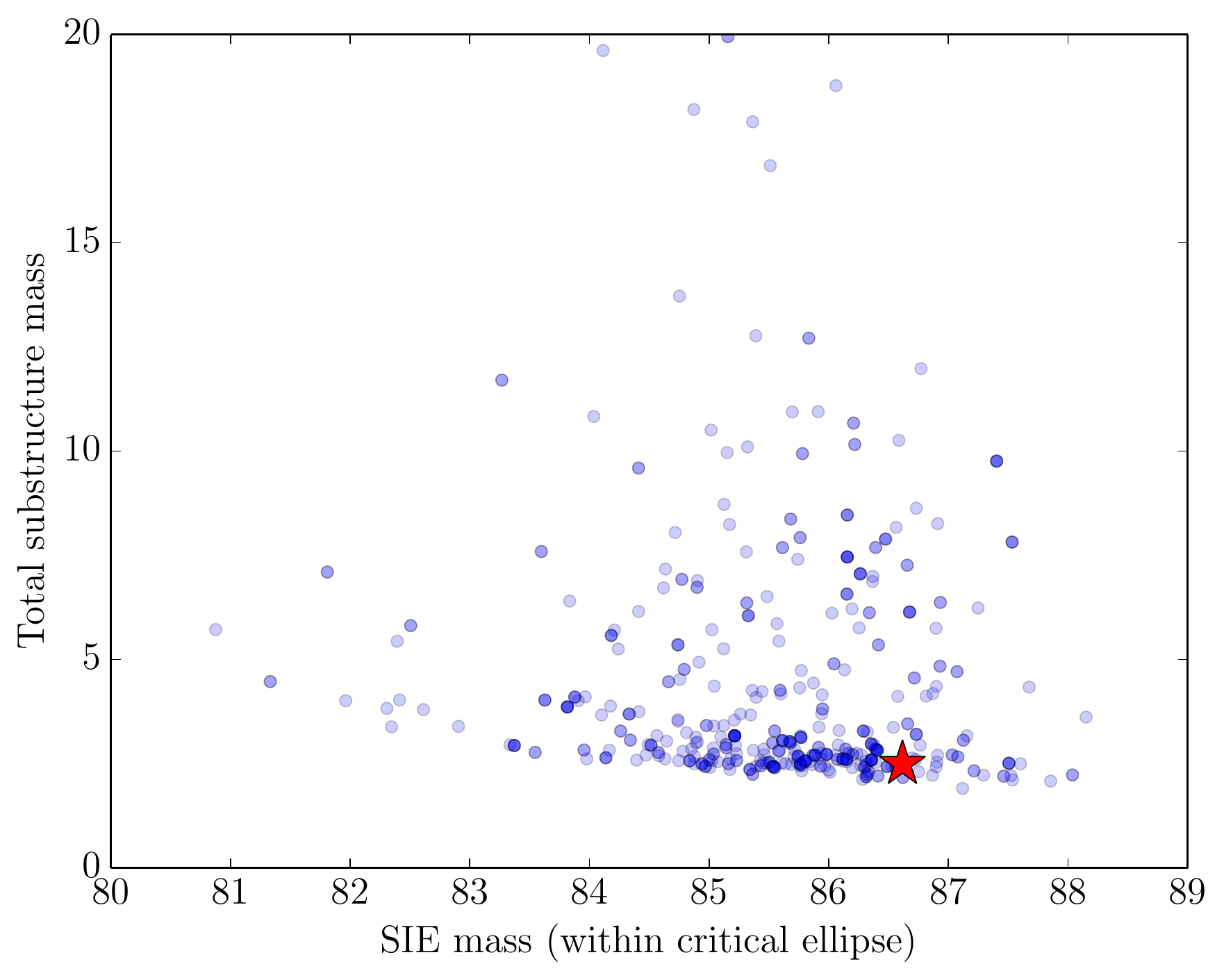}
\caption{The joint posterior distribution for the SIE mass (integrated within
its critical ellipse, which is not the critical curve of the lens overall),
and the total mass in substructures.\label{fig:masses1}}
\end{center}
\end{figure}

In the present paper we do not address the question of exactly which quantities
related to the
density profile of the lens are most scientifically interesting. Nevertheless,
we can verify that the results of the inference do behave in understandable
ways. For example, in Figure~\ref{fig:masses1},
we plot the joint posterior distribution
for the SIE mass within its critical ellipse and the total substructure mass
over the entire domain. As one would expect, there is a strong dependence
between these two quantities in the posterior distribution, as mass in the
SIE can be traded off with mass in substructures to some extent, while
the model remains (loosely speaking) ``consistent with the data''.

\begin{figure}
\begin{center}
\includegraphics[scale=0.4]{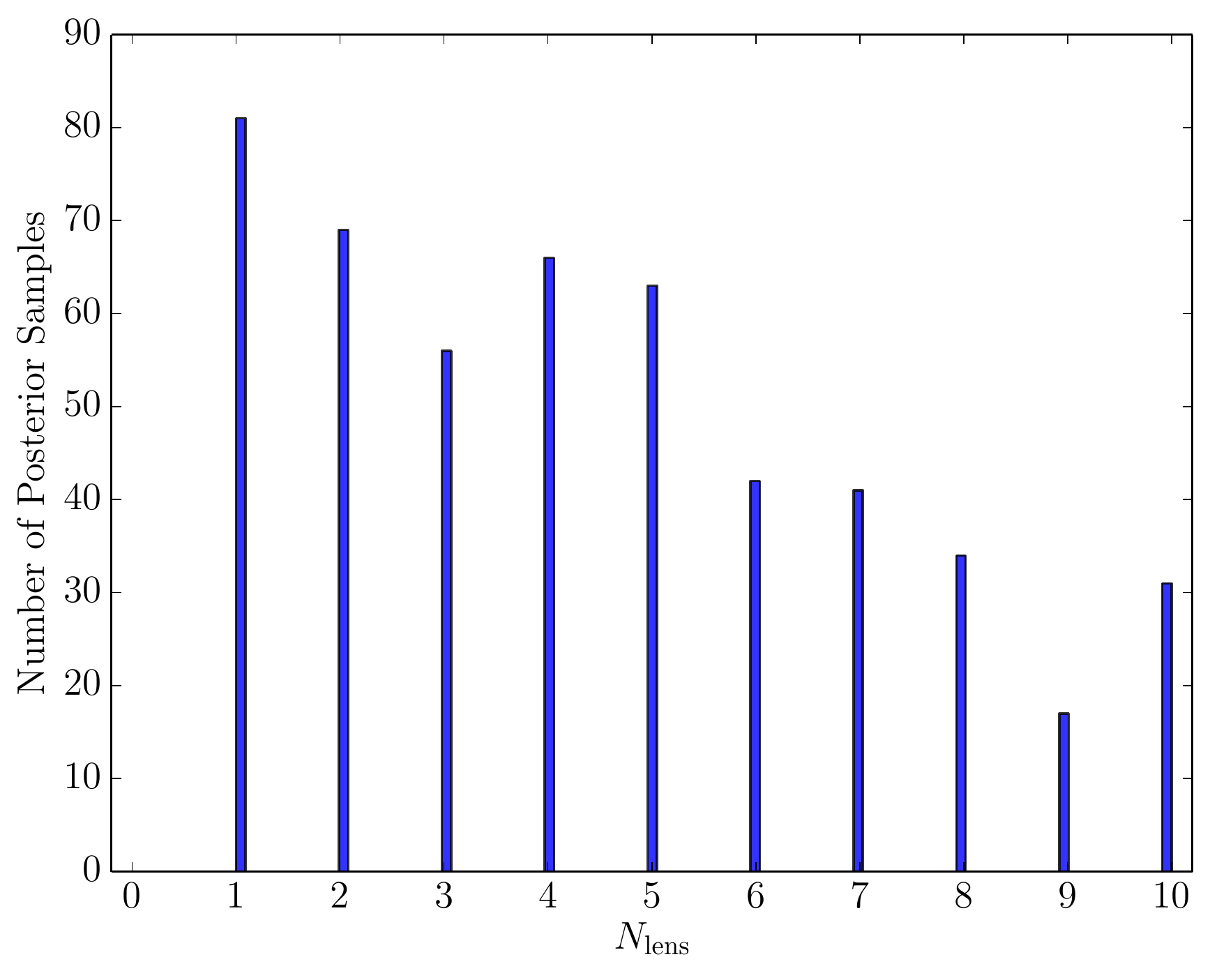}
\caption{The marginal posterior distribution for $N_{\rm lens}$, the
number of substructures in the lens. The prior was uniform and the true
value used to generate the data was 1. However, the data is only informative
enough to suppress the probability of values above 1 slightly, since it
is possible (given this data) that low mass substructures might exist somewhere,
or that what we think is one substructure might actually be two close together,
and other such possibilities.
\label{fig:N_lens1}}
\end{center}
\end{figure}

The posterior distribution for $N_{\rm lens}$ is also clearly of interest, and
is displayed in Figure~\ref{fig:N_lens1}. The prior for this parameter was
uniform from 0 to 10 (inclusive), and the possibility $N=0$ has been
(loosely speaking) ``ruled out'' by the image data. The true solution ($N=1$)
has the highest probability. However, the possibilities with $N > 1$ are
still fairly plausible, since it's possible that a very small substructure
exists, and it's also possible that two substructures near each other could
mimic the effect of one. The degree to which these possibilities are plausible
is related to the choice of prior for the blob amplitudes and positions.

The estimated marginal likelihood of our model
(averaged over all parameters including $N_{\rm lens}$ and $N_{\rm src}$)
is {\revisions
$\ln\left[p(\boldsymbol{D})\right] \approx -14141.4$, and
the information (Kullback-Leibler divergence from the prior to the posterior)
is $\mathcal{H} \approx 99.4$ nats}. The information represents the degree of
compression of the posterior distribution with respect to the prior, and can
be interpreted quite literally as how much was learned about the parameters from
the data. It is also straightforward to estimate from Nested Sampling
\citep{skilling}. Its definition is:
\begin{eqnarray}
\mathcal{H} = \int p(\boldsymbol{\theta} | \boldsymbol{D})
\ln\left[\frac{p(\boldsymbol{\theta} | \boldsymbol{D})}{p(\boldsymbol{\theta})}\right]
\, d\boldsymbol{\theta}
\end{eqnarray}
and we can build intuition about its meaning based on some simple examples. One
such example is a uniform prior over a volume $V_0$ and a posterior which is
uniform over a smaller volume $V_1$ contained within $V_0$. In this case
$\mathcal{H} = \ln(V_0/V_1)$ nats. Therefore, a value of $\mathcal{H}=100$ nats
implies the posterior distribution occupies roughly $e^{-100}$ of the prior
volume.

\begin{figure}
\begin{center}
\includegraphics[scale=0.4]{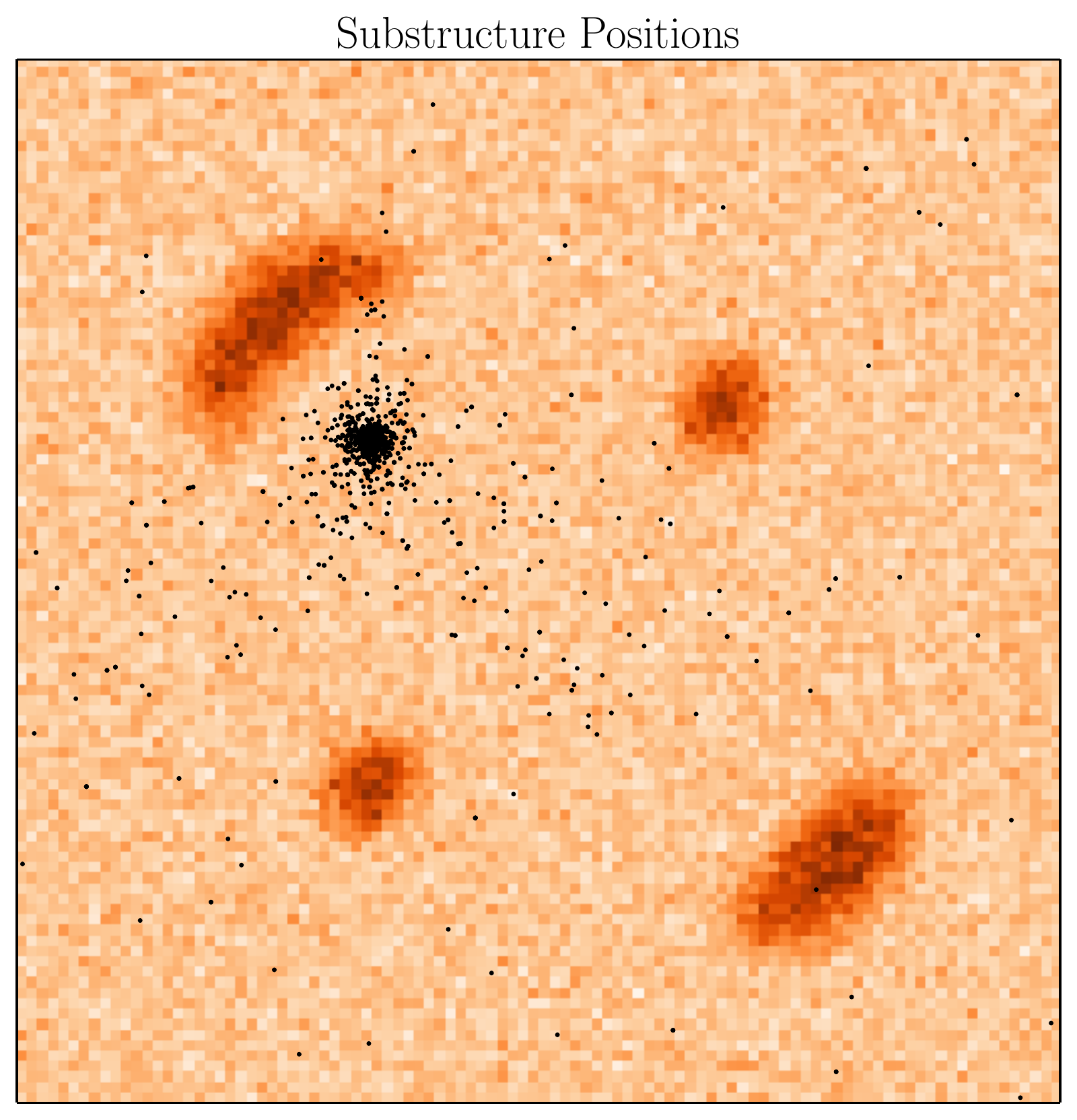}
\caption{The {\revisions `easy'} simulated data, with substructure positions overlaid (a
Monte Carlo approximation to the posterior expectation of the empirical
measure). The density of black points in any region is proportional to the
expected number of substructures whose centers lie within that region. In this
case, there is strong evidence for a substructure close to the top-left image
(where one was actually placed).
\label{fig:substructures1}}
\end{center}
\end{figure}

{\revisions
For the sampled parameter values representing the posterior distribution, there
was no structure in the residuals (not shown), when normalised by the noise
standard deviation in each pixel. This is unsurprising since the model
assumptions are entirely correct for this dataset.
}

\begin{figure}
\begin{center}
\includegraphics[scale=0.5]{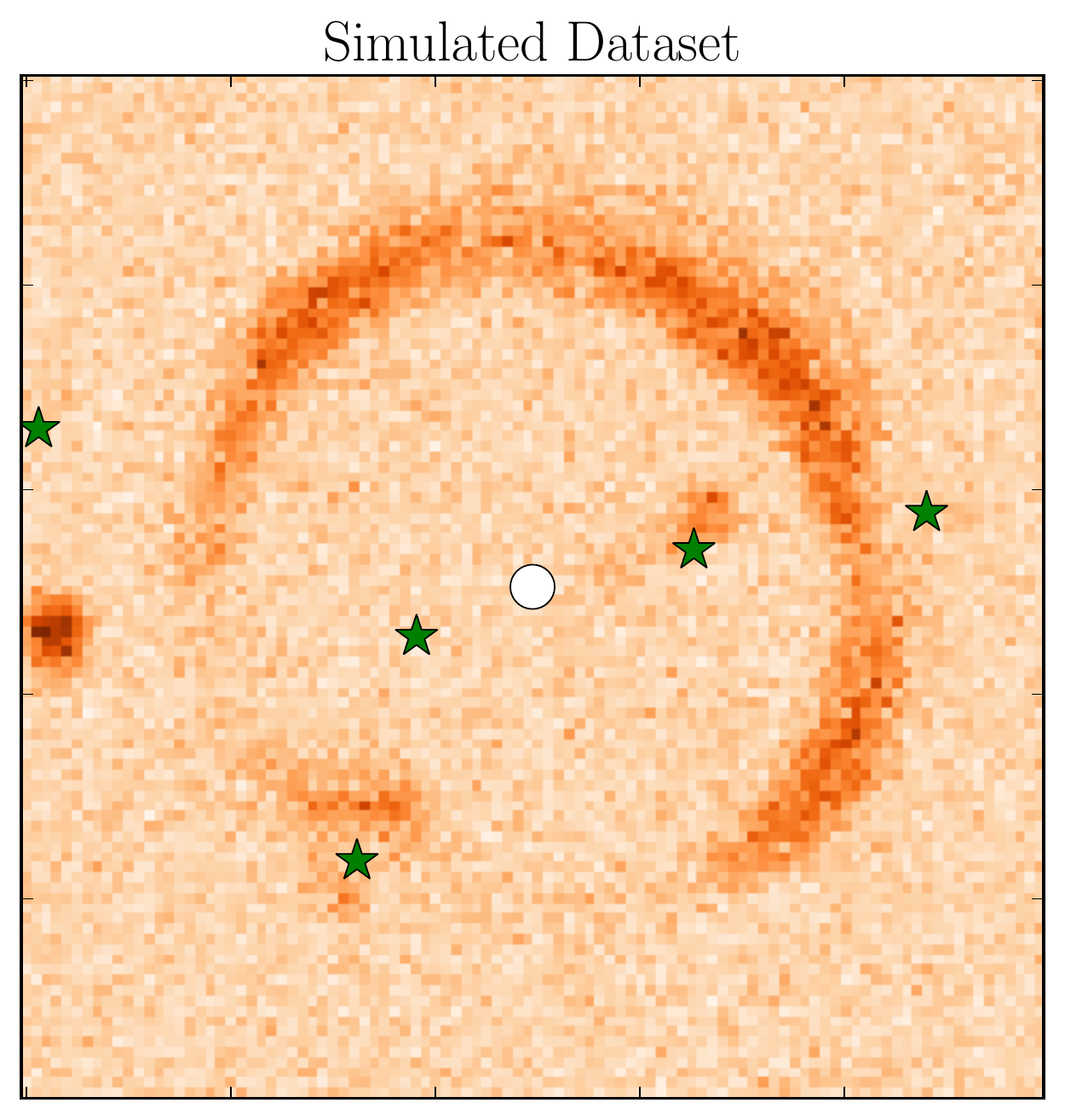}
\caption{{\revisions The `harder' simulated data, consisting of a more
complex source profile, and a lens with an SIE+$\gamma$ (whose centre is
indicated by a white circle) and ten substructures. Only five of the
substructures (star symbols) occured within the square image boundary, but the other five still have an effect on the image.}
\label{fig:image2}}
\end{center}
\end{figure}

{\revisions
\section{``Harder'' Simulated Data}
To further test the algorithm, we created a ``harder'' simulated image, where
the lens consisted of an SIE+$\gamma$ profile plus 10 additional substructures.
The image is shown in Figure~\ref{fig:image2}. With the easy dataset, there
was little hope of measuring the substructure mass function parameter
$\mu_{\rm lens}$ since there was only a single substructure which would provide
little information about $\mu_{\rm lens}$ (only constraining its general
order of magnitude), even if its mass were measured perfectly.}

\begin{figure}
\begin{center}
\includegraphics[scale=0.4]{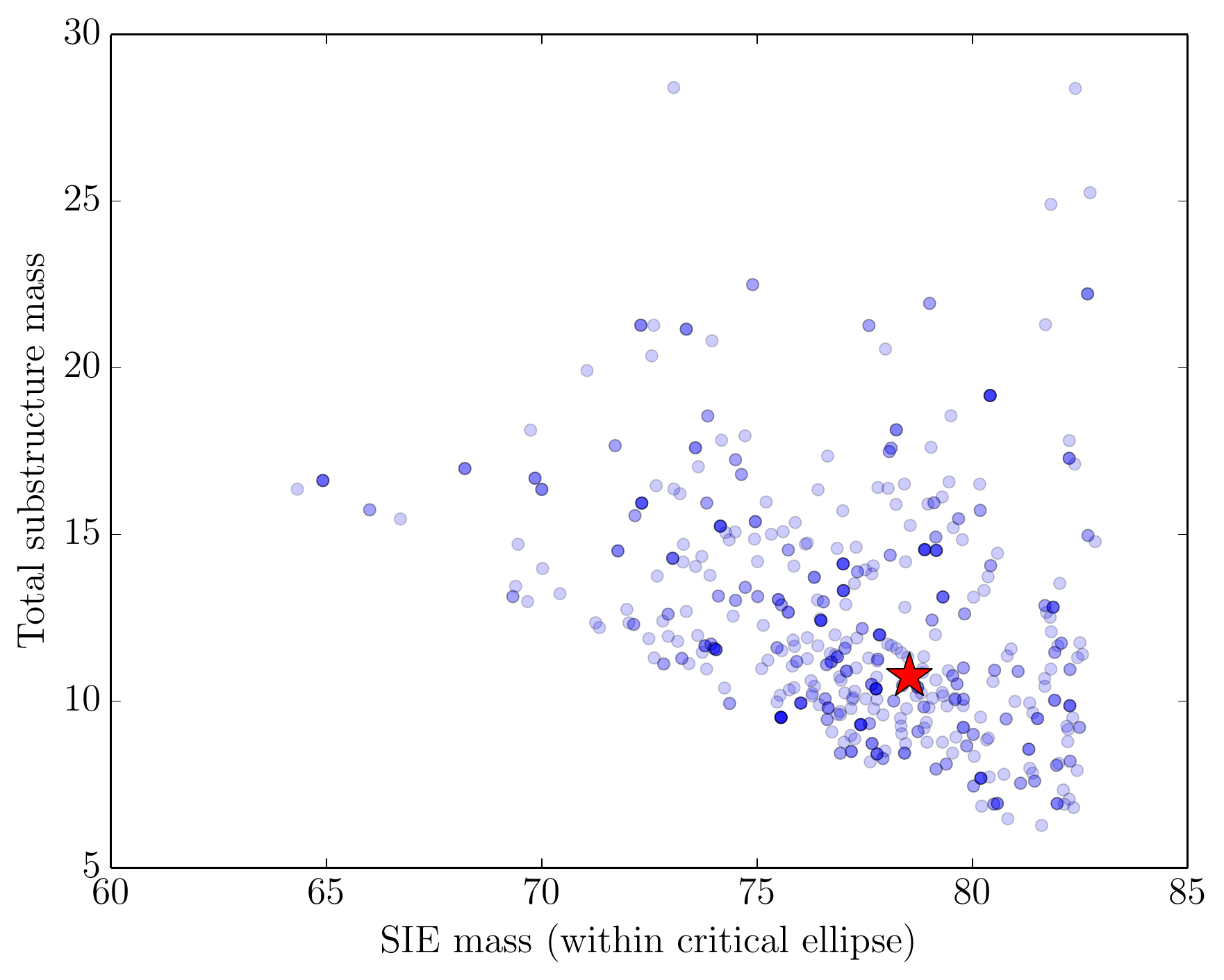}
\caption{{\revisions
Same as Figure~\ref{fig:masses1}, but for the `harder' simulated dataset.
}
\label{fig:masses2}}
\end{center}
\end{figure}

{\revisions As with the `easy' dataset,
we generated 5,000 samples from the DNS target distribution and thinned
by a factor of $10^5$, so $5 \times 10^8$ MCMC iterations were actually
performed, taking approximately 60 hours. The slower runtime in this case
was because $N_{\rm lens} = 10$, more CPU time was spent exploring parts of
the hypothesis space where $N_{\rm lens}$ is high. The run resulted in 557
equally-weighted posterior samples. The posterior distribution for the SIE
mass and the total substructure mass is given in Figure~\ref{fig:masses2}.
The samples can be viewed as a movie at
{\tt www.youtube.com/watch?v=o3ppfKSk248}.
}

\begin{figure}
\centering
\includegraphics[scale=0.45]{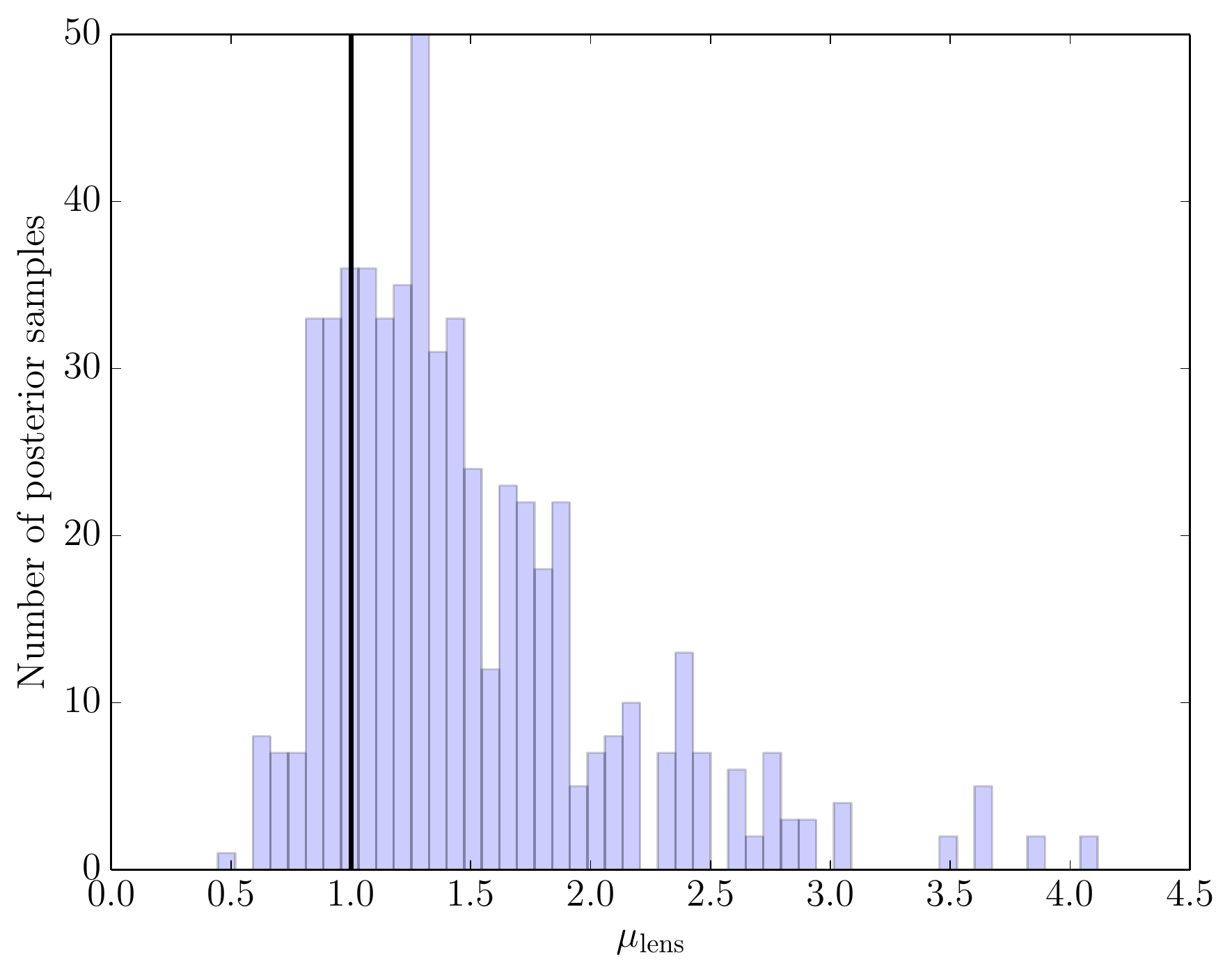}
\caption{{\revisions The posterior distribution for $\mu_{\rm lens}$, the
hyperparameter which determines the prior expected value of substructure
masses, given the `harder' simulated data. The true value (in arbitrary units)
was 1, which is indicated by the vertical line.
}
\label{fig:mu_lens}}
\end{figure}

{\revisions
To demonstrate the claim that we can infer something about the mass function
of substructures directly from the image data, we have plotted the posterior
distribution for $\mu_{\rm lens}$, the hyperparameter for the substructure
masses, in Figure~\ref{fig:mu_lens}. The true value was 1, and the
posterior distribution indicates a fairly wide uncertainty range that is
at least consistent with the true value in some sense. Given that the image
only contained a few substructures whose masses could be measured with any
accuracy, the large uncertainty is not surprising.
}

\begin{figure}
\begin{center}
\includegraphics[scale=0.4]{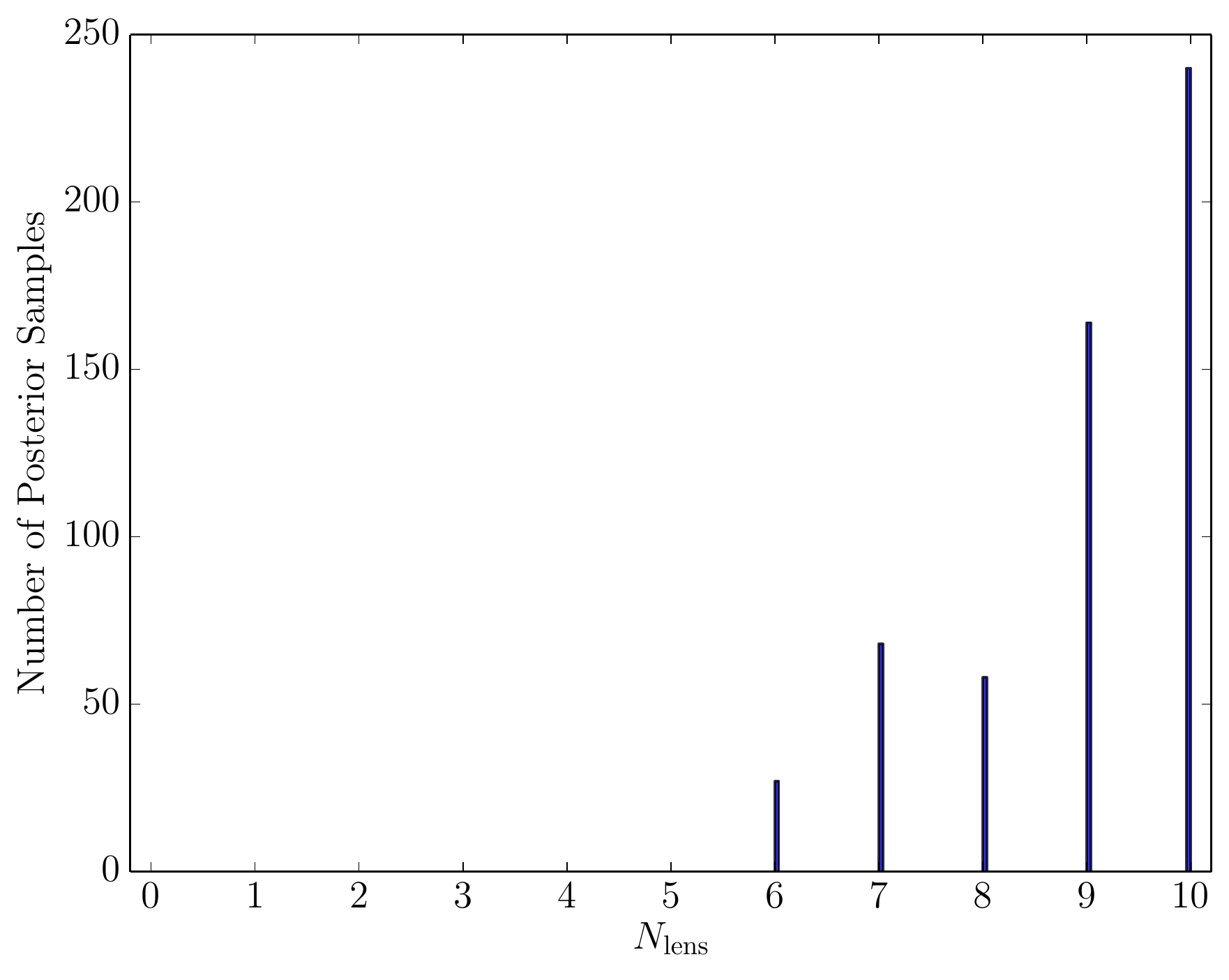}
\caption{{\revisions
Same as Figure~\ref{fig:N_lens1}, but for the `harder' simulated dataset.
The true number of substructures was 10, which also happens to be the posterior
mode. Only five substructures were located within the domain of the image, but
$N_{\rm lens} = 10$ is probable because the spread-out positions of the
substructures imply that $R_{\rm lens}$ is probably large.
}
\label{fig:N_lens2}}
\end{center}
\end{figure}

\begin{figure}
\begin{center}
\includegraphics[scale=0.4]{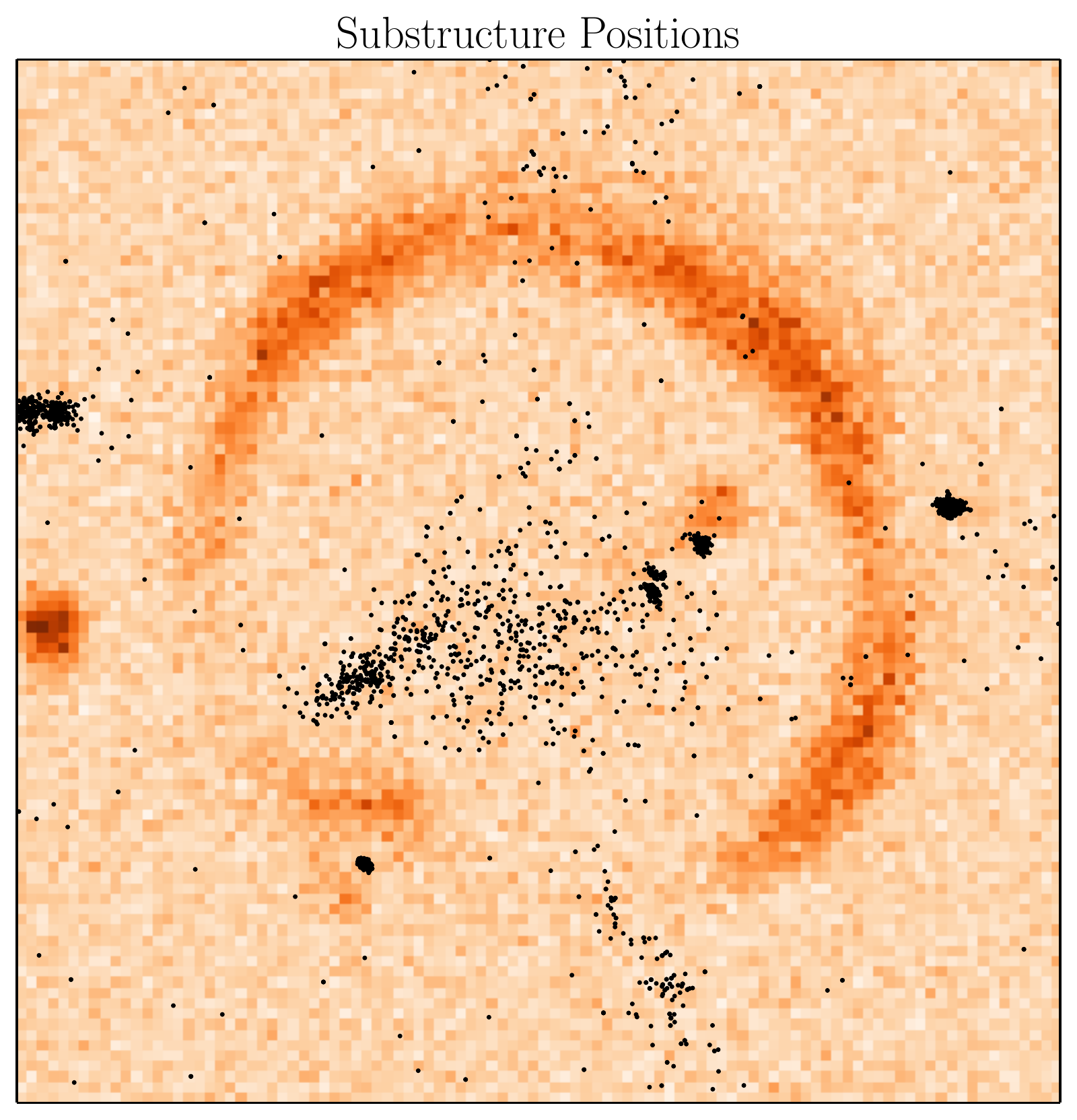}
\caption{{\revisions
Same as Figure~\ref{fig:substructures1}, but for the `harder' simulated dataset.
For all of the substructures located close to the ring, the positions were well
constrained. There was weak evidence for some substructures that did not in
fact exist.
}
\label{fig:substructures2}}
\end{center}
\end{figure}

{\revisions
The estimated marginal likelihood of our model, for the `harder' simulated data,
is
$\ln\left[p(\boldsymbol{D})\right] \approx -5671.9$, and
the information (Kullback-Leibler divergence from the prior to the posterior)
is $\mathcal{H} \approx 155.7$ nats. 
As with the easy dataset, the standardised residuals of the sampled models
resembled noise.
}

\section{The Cosmic Horseshoe}
\begin{figure}
\begin{center}
\includegraphics[scale=0.5]{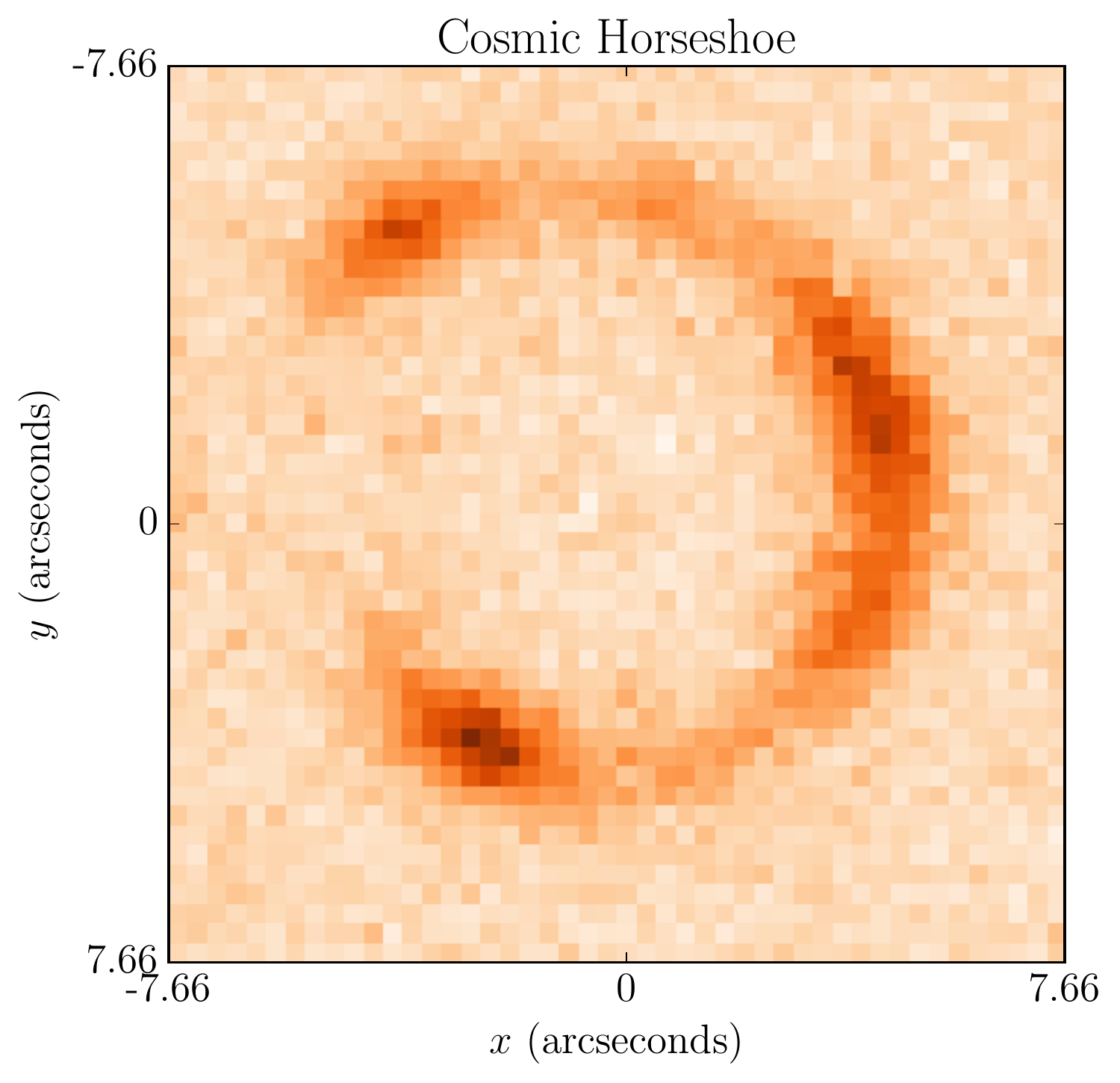}
\caption{The g-band INT image of the Cosmic Horseshoe, with the lens galaxy
subtracted.
\label{fig:image3}}
\end{center}
\end{figure}

As a further demonstration the method, we apply it to the g-band data
of the Cosmic Horseshoe J1004+4112 \citep{belokurov, 2008MNRAS.388..384D} taken with the Isaac Newton Telescope (INT). The image is shown in Figure~\ref{fig:image3}.
For this system,
more data is available (i and U band data, as well as more recent
Hubble Space Telescope (HST) imaging), although the g-band image is the highest
signal-to-noise of the INT images. The source redshift is $z_s=2.379$, and the
lens is a massive luminous red galaxy at $z_l=0.4457$.
Unfortunately our current implementation
doesn't allow for multi-band data (unlike \citet{2011MNRAS.412.2521B}), and
is quite slow when running on the larger HST image. Therefore this section
should be considered a further demonstration of the technique, and not a
thorough study of this system.

We generated 5,000 samples from the DNS target distribution and thinned
by a factor of $10^5$, so $5 \times 10^8$ MCMC iterations were actually
performed, taking approximately 40 hours. After resampling, this resulted in
{\revisions 718}
equally weighted posterior samples. Although this image is smaller than the
simulated data ($46 \times 46$ pixels), we fired $2 \times 2$ rays per pixel
for greater accuracy.

\begin{figure}
\begin{center}
\includegraphics[scale=0.4]{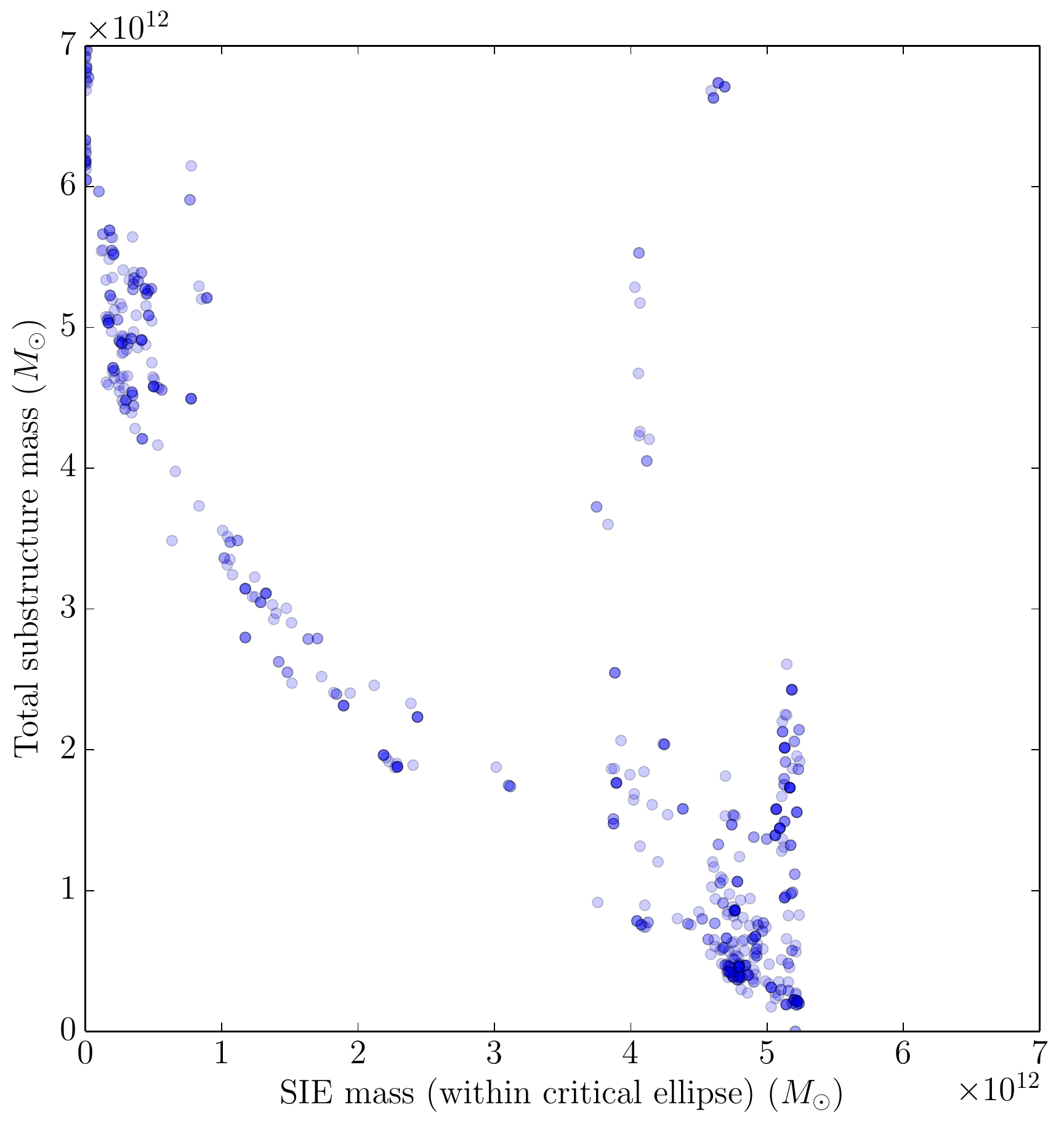}
\caption{{\revisions Same as Figure~\ref{fig:masses1}, but for the Cosmic Horseshoe.} The joint posterior distribution for the SIE mass (integrated within
its critical ellipse, which is not the critical curve of the lens overall),
and the total mass in substructures. The units are defined by the critical
density, so that a mass of $\pi$ units would have an Einstein radius of one arcsecond. To calculate the masses in solar masses, we assumed a flat cosmology with
$\Omega_m=0.3$, $\Omega_\Lambda=0.7$, and
$H_0=70$ km/s/Mpc.
\label{fig:masses3}}
\end{center}
\end{figure}

The joint posterior distribution for the SIE mass total substructure mass
for the Cosmic Horseshoe is given in Figure~\ref{fig:masses3}. As with
the simulated data (Figure~\ref{fig:masses1}), we see an expected negative
correlation between these two quantities. However, the points are not as
smoothly distributed in this case (we discuss this issue further
in Section~\ref{sec:convergence}). An upper ``limit'' on the SIE mass, with
95\% posterior probability, is $5.20 \times 10^{12}$ solar masses.
The algorithm has also found
some possibilities where the substructure mass is much greater than this. In
these cases, the substructures are far from the image --- it is the environment
that is being modelled.

\begin{figure}
\begin{center}
\includegraphics[scale=0.4]{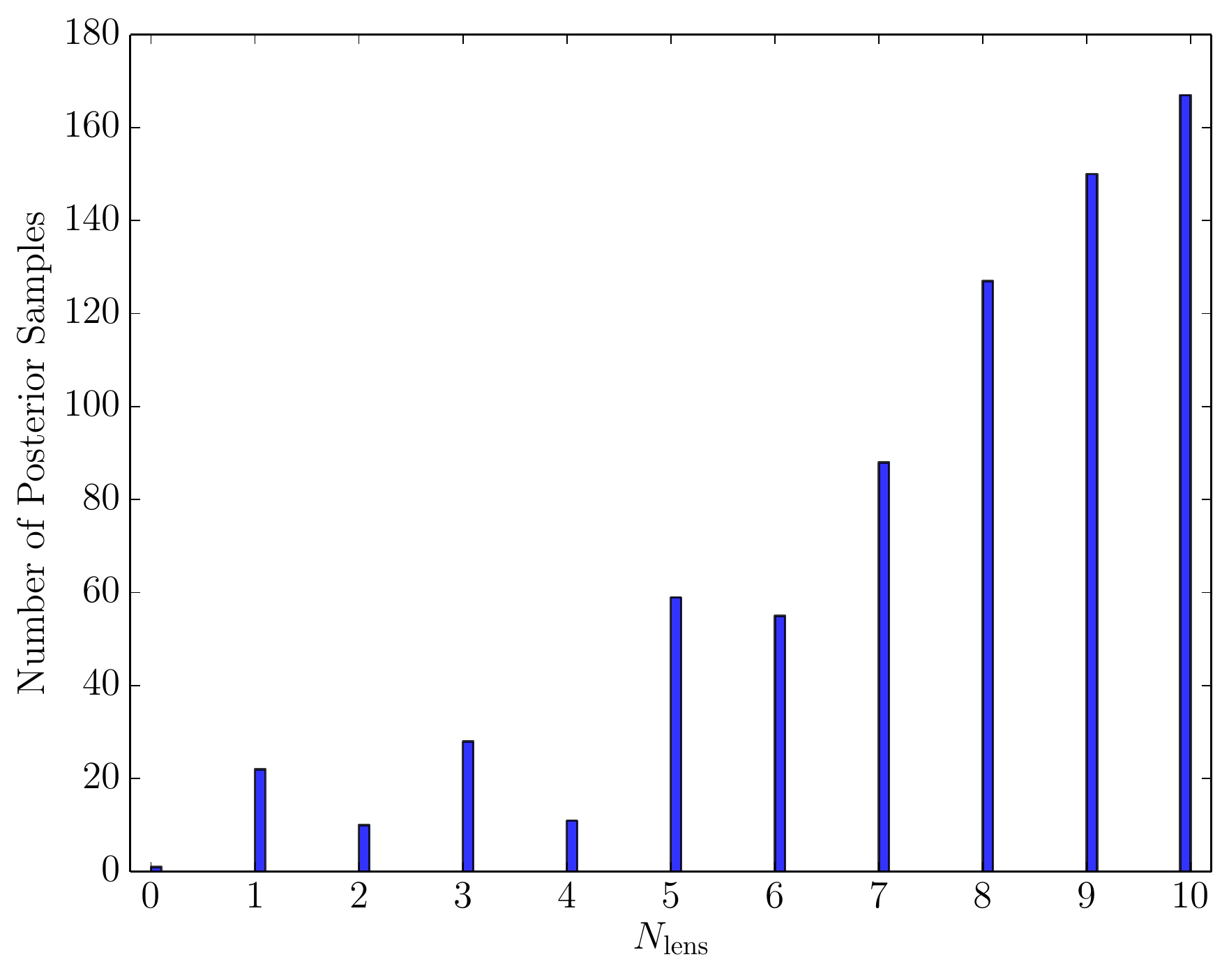}
\caption{{\revisions Same as Figure~\ref{fig:N_lens1}, but for the Cosmic Horseshoe.} The posterior distribution for the number of lens substructures
in the Cosmic Horseshoe system. There is {\revisions moderate} evidence in favour of the hypothesis that $N_{\rm lens} \neq 0$.\label{fig:N_lens3}}
\end{center}
\end{figure}

The posterior distribution for
$N_{\rm lens}$ (Figure~\ref{fig:N_lens3})
also shows some evidence for more than zero substructures.
In particular, the prior probability for $N_{\rm lens} > 0$ was 10/11
$\approx$ 0.91, whereas the posterior probability is approximately {\revisions 0.997}.
This corresponds to a Bayes Factor of {\revisions about 300} in favour of
$N_{\rm lens} > 0$ versus the alternative $N_{\rm lens} = 0$.

\begin{figure}
\begin{center}
\includegraphics[scale=0.4]{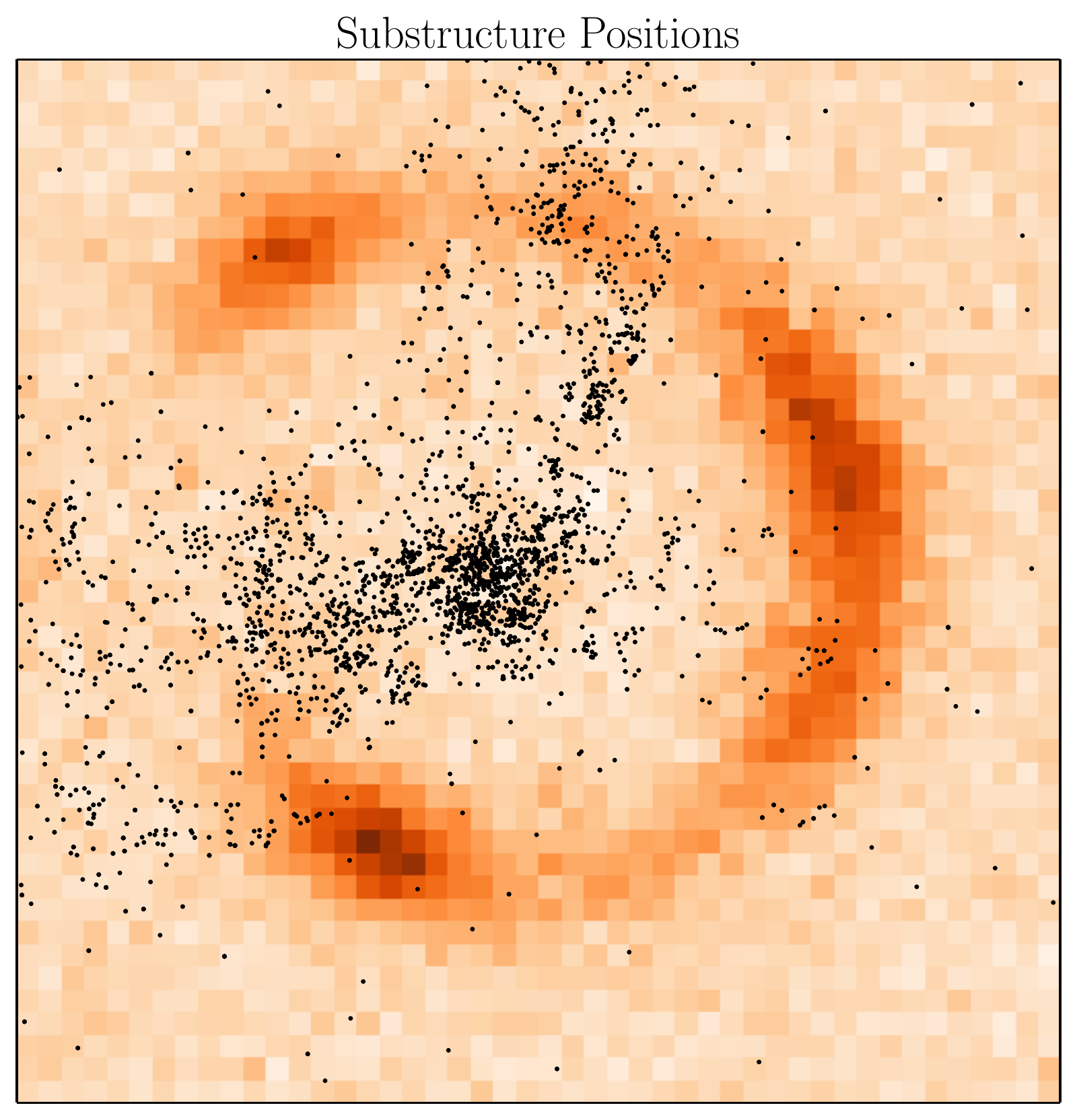}
\caption{{\revisions Same as Figure~\ref{fig:substructures1}, but for the Cosmic Horseshoe.} The positions of substructures encountered in the MCMC sampling
for the Cosmic Horseshoe; technically a Monte Carlo approximation to the
posterior expected value of the empirical measure of substructure
positions. Unlike Figure~\ref{fig:substructures1}, for the Cosmic Horseshoe
there is little consistency in the positions of the substructures.
\label{fig:substructures3}}
\end{center}
\end{figure}

Despite weak evidence for the existence of substructure, when we examine the
expected value of the empirical measure of substructure positions
(Figure~\ref{fig:substructures3}) we find no consistency in their positions,
unlike for the simulated data (Figure~\ref{fig:substructures1}). This situation
is not uncommon. For example, \citet{exoplanet} found evidence for a large
number of Keplerian signals in a time series, but the number of such signals
with well constrained properties was much lower. Related to this, we can
compute posterior probabilities for any hypothesis about the substructure
masses. With 95\% probability the substructure mass is less than {\revisions 1.53$\times 10^{13}$} solar masses and with 75\% probability it is less than {\revisions 5.10$\times 10^{12}$} solar masses. As the Einstein ring itself implies a mass of about $5 \times 10^{12}$
solar masses, these summaries are affected by the possibility of blobs outside
the main Einstein ring.

As with any inference, the results presented here may be sensitive to many of
the input modelling assumptions, and a slightly different (yet still
reasonable) set of choices might yield different results. For example, if
the density profile of the lens was smooth but not of the SIE form, and the
data were informative enough to show this, the current
model would only be able to explain the data by adding substructures. Hence, an alternative explanation for these results is that, rather than containing a substructure, the density profile of the lens is simply not within the SIE$+\gamma$ family. In fact, some of the posterior
samples obtained contain massive substructures outside of the image, which
could be modelling higher-order effects of the environment beyond what is
captured by the simple external shear model. This is one way of violating
the SIE$+\gamma$ assumption, and another is simply to have a different projected
density profile.
One way of further investigating this is to use a different family of
smooth lens models and doing model selection based on the marginal likelihoods.
Another option which we defer to future work is to implement a more flexible
model (such as a mixture of concentric elliptical lenses). This
result is consistent with that of \citet{2008MNRAS.388..384D}, who found a slight preference for an elliptical power law profile over the SIE
(which is a specific case of an elliptical power law).

Another potential inadequacy of our model is the assumption that the PSF is
known. In practice, the PSF is often estimated from the image of nearby star(s),
or from theoretical knowledge of the telescope (especially in the case of
space based imaging). However, if the PSF is misspecified, or in fact varies
across the image, this could induce subtle effects in the imaging which our
current model could only explain using substructure.

\begin{figure*}
\begin{center}
\includegraphics[scale=0.8]{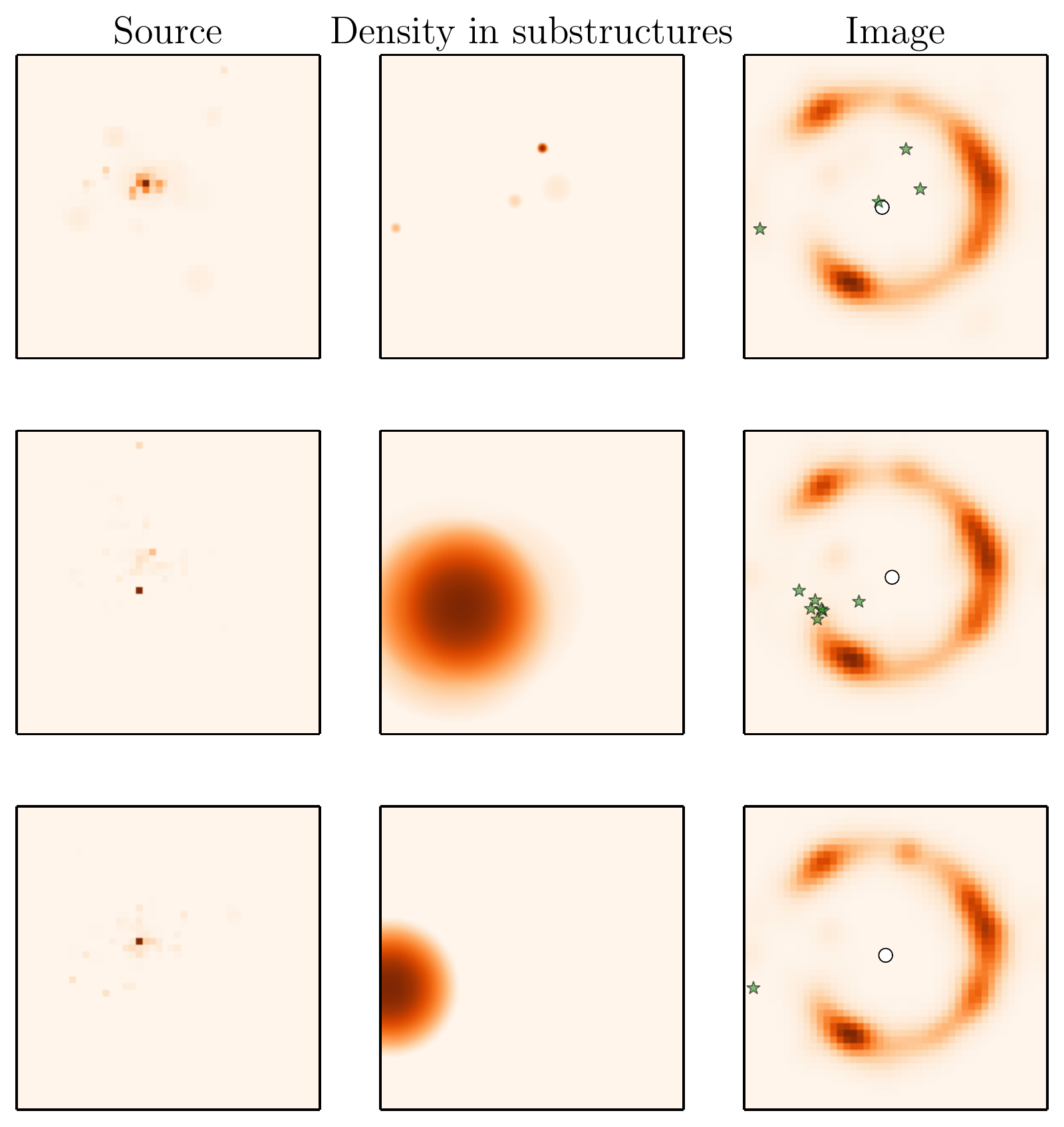}
\caption{Three example sources, the density profile in substructures,
and the corresponding lensed, blurred images,
representative of the posterior distribution.
The angular scales are 7.5 arcseconds for the sources and 15
arcseconds for the images.
\label{fig:sources}}
\end{center}
\end{figure*}

The estimated marginal likelihood of our model
is {\revisions
$\ln\left[p(\boldsymbol{D})\right] \approx -6948.5$}, and
the information (Kullback-Leibler divergence from the prior to the posterior)
is {\revisions $\mathcal{H} \approx 123.1$} nats. 
Three example models (lens and source) sampled from the posterior are shown
in Figure~\ref{fig:sources}. {\revisions
The model was able to fit the data down to the noise level.
}

\subsection{Reproducibility of the results}\label{sec:convergence}
The ``effective sample size'' returned by DNS, which we have described as the
number of posterior samples, takes into account the fact DNS's target
distribution is not the posterior. However, it does not take autocorrelation
into account, and thus can present an optimistic picture of the accuracy of
any Monte Carlo approximations to posterior quantities. Most standard diagnostic
techniques used in MCMC can be applied here. The simplest of these is a
check of reproducibility. If different runs yield substantially different
results, the MCMC output should be treated with caution. For example, in
Figure~\ref{fig:masses3}, there is a correlation between the mass attributed
to the SIE component and that attributed to substructures. However, the
distribution also appears somewhat ``lumpy'' or multimodal. This may
not be a feature of the actual posterior distribution, but could arise due to
imperfect sampling. Whereas a standard Metropolis sampler would move around
very slowly in the hypothesis space, DNS naturally spends a non-negligible fraction
of the time sampling the prior. Therefore, a particle exploring the parameter
space can ``forget'' its good-fitting position, move somewhere completely
different, and find another good-fitting model in a different location. This
is a natural feature of the algorithm (and is also present in related
algorithms such as parallel tempering). Despite this, the patchy
nature of Figure~\ref{fig:masses3} suggests that posterior exploration is
still challenging in this problem.

The results for the simulated data (Figure~\ref{fig:masses1})
were less complex because of the definite existence of one substructure.
Its mass and position provide some information about the hyperparameters
$\boldsymbol{\alpha}_{\rm lens}$, restricting the probability that
high mass substructures exist far from the image.

\section{Conclusions}
We have developed a trans-dimensional Bayesian approach motivated directly by
the question of whether dark substructures exist in a lens galaxy, given image
data. By making use of the Diffusive Nested Sampling algorithm \citep{dnest}
and the {\tt RJObject} library \citep{rjobject}, we outsource the difficulties
associated with choosing Metropolis proposals for a hierarchical model of
non-fixed dimension. The model allows for source and lens ``blobs'' to appear
as needed to explain the data. The prior for the blobs' properties is
specified hierarchically, to more realistically model sensible prior beliefs,
and to tie the model parameters directly to questions of scientific interest
such as the mass function of substructures.

As a proof of concept, we demonstrated the successful recovery of a single
substructure from simulated data for which all of the model assumptions were
true. We then applied the method to an image of the Cosmic Horseshoe system
and found {\revisions moderate} evidence in favour of the existence of a substructure, or at
the very least, a departure from a simple SIE+$\gamma$ lens profile. The
main output of our method is a set of posterior samples, each representing
a plausible scenario for the lens and the source given the data
and the assumed prior information. These samples can be displayed as a movie,
which is a very useful method for intuitively understanding the remaining
uncertainty. Any posterior summaries of interest can be
produced from the samples, but the interpretation of many of these summaries
is not necessarily straightforward. We have suggested that the
posterior expectation of the empirical measure of substructure positions is
one of the most helpful summaries.

The model assumptions used in this paper are fairly simple. In future work we
intend to generalize the model to multi-band data, and extend the
smooth lens model beyond the overly simplistic
SIE+$\gamma$ assumption. Applications
to other systems are also forthcoming, as well as variants on the model using
different conditional priors for the substructure properties given the
hyperparameters.

\section*{Acknowledgements}
It is a pleasure to thank Phil Marshall (Stanford), Tommaso Treu (UCLA),
Ross Fadely (NYU), Alan Heavens (Edinburgh), {\revisions and the referee}
for valuable discussion. Thomas Lumley (Auckland) also deserves credit for
ending BJB's usage of the jet colormap.
This work was funded by a Marsden Fast Start grant from the Royal Society of
New Zealand.


\begin{thebibliography}{99}
\bibitem[\protect\citeauthoryear{Alsing et al.}{2015}]{2015arXiv150507840A} 
Alsing J., Heavens A., Jaffe A.~H., Kiessling A., Wandelt B., Hoffmann T., 
2015, arXiv, arXiv:1505.07840

\bibitem[\protect\citeauthoryear{Auger et al.}{2011}]{eels} 
Auger M.~W., Treu T., Brewer B.~J., Marshall P.~J., 2011, MNRAS, 411, L6

\bibitem[\protect\citeauthoryear{Barkana}{1998}]{fastell} 
Barkana R., 1998, ApJ, 502, 531

\bibitem[\protect\citeauthoryear{Belokurov et 
al.}{2007}]{belokurov} Belokurov V., et al., 2007, ApJ, 671, L9

\bibitem[\protect\citeauthoryear{Birrer, Amara, 
\& Refregier}{2015}]{2015arXiv150407629B} Birrer S., Amara A., Refregier A., 2015, arXiv, arXiv:1504.07629

\bibitem[\protect\citeauthoryear{Brewer 
\& Lewis}{2006}]{2006ApJ...637..608B} Brewer B.~J., Lewis G.~F., 2006, ApJ, 637, 608

\bibitem[\protect\citeauthoryear{Brewer, P{\'a}rtay,
\& Cs{\'a}nyi}{2011}]{dnest} Brewer B.~J., P{\'a}rtay L.~B., Cs{\'a}nyi G., 2011,
Statistics and Computing, 21, 4, 649-656. arXiv:0912.2380

\bibitem[\protect\citeauthoryear{Brewer et al.}{2011}]{2011MNRAS.412.2521B} 
Brewer B.~J., Lewis G.~F., Belokurov V., Irwin M.~J., Bridges T.~J., Evans 
N.~W., 2011, MNRAS, 412, 2521

\bibitem[\protect\citeauthoryear{Brewer}{2014}]{rjobject} Brewer, B. J., 2014,
preprint. ArXiv: 1411.3921

\bibitem[\protect\citeauthoryear{Brewer 
\& Donovan}{2015}]{exoplanet} Brewer B.~J., Donovan C.~P., 2015, MNRAS, 448, 3206 

\bibitem[Caticha(2008)]{caticha} Caticha, A.\ 2008.\ Lectures 
on Probability, Entropy, and Statistical Physics.\ ArXiv e-prints 
arXiv:0808.0012. 

\bibitem[\protect\citeauthoryear{Coles, Read, 
\& Saha}{2014}]{2014MNRAS.445.2181C} Coles J.~P., Read J.~I., Saha P., 2014, MNRAS, 445, 2181

{\revisions
\bibitem[\protect\citeauthoryear{Dehnen 
\& Aly}{2012}]{2012MNRAS.425.1068D} Dehnen W., Aly H., 2012, MNRAS, 425, 1068
}

\bibitem[\protect\citeauthoryear{Dye et al.}{2008}]{2008MNRAS.388..384D} 
Dye S., Evans N.~W., Belokurov V., Warren S.~J., Hewett P., 2008, MNRAS, 
388, 384 

\bibitem[\protect\citeauthoryear{Fadely 
\& Keeton}{2012}]{2012MNRAS.419..936F} Fadely R., Keeton C.~R., 2012, MNRAS, 419, 936

\bibitem[\protect\citeauthoryear{Green}{1995}]{green}
Green, P.~J., 1995, Reversible Jump Markov Chain Monte Carlo Computation and Bayesian Model Determination, Biometrika 82 (4): 711–732.

\bibitem[\protect\citeauthoryear{Grillo et al.}{2015}]{grillo} 
Grillo C., et al., 2015, ApJ, 800, 38 

\bibitem[\protect\citeauthoryear{Huppenkothen et 
al.}{2015}]{magnetron} Huppenkothen D., et al., 2015, ApJ, 810, 
66 


\bibitem[\protect\citeauthoryear{Jones, Kashyap, 
\& van Dyk}{2014}]{jones} Jones D.~E., Kashyap V.~L., van Dyk D.~A., 2014, arXiv, arXiv:1411.7447

\bibitem[\protect\citeauthoryear{Koopmans}{2005}]{koopmans} 
Koopmans L.~V.~E., 2005, MNRAS, 363, 1136

\bibitem[\protect\citeauthoryear{Kormann, Schneider, 
\& Bartelmann}{1994}]{1994A&A...284..285K} Kormann R., Schneider P., Bartelmann M., 1994, A\&A, 284, 285

\bibitem[\protect\citeauthoryear{Millar}{2011}]{millar}
Millar, R.~B., Maximum likelihood estimation and inference: with examples in R, SAS and ADMB. Vol. 111. John Wiley \& Sons, 2011.

\bibitem[\protect\citeauthoryear{O'Hagan and Forster}{2004}]{ohagan}
O'Hagan, A., Forster,~J., 2004, Bayesian inference. London: Arnold.

\bibitem[\protect\citeauthoryear{Pancoast, Brewer, 
\& Treu}{2011}]{pancoast} Pancoast A., Brewer B.~J., Treu T., 2011, ApJ, 730, 139

\bibitem[\protect\citeauthoryear{Schneider et 
al.}{2015}]{2015ApJ...807...87S} Schneider M.~D., Hogg D.~W., Marshall 
P.~J., Dawson W.~A., Meyers J., Bard D.~J., Lang D., 2015, ApJ, 807, 87

\bibitem[\protect\citeauthoryear{Sivia \& Skilling}{2006}]{sivia} Sivia, 
D.~ S., Skilling, J., 2006, Data Analysis: A Bayesian Tutorial, 2nd 
Edition, Oxford University Press

\bibitem[\protect\citeauthoryear{Skilling}{2006}]{skilling} Skilling, 
J., 2006, ``Nested Sampling for General Bayesian Computation'', Bayesian 
Analysis 4, pp. 833-860

\bibitem[\protect\citeauthoryear{Suyu et al.}{2006}]{suyu} 
Suyu S.~H., Marshall P.~J., Hobson M.~P., Blandford R.~D., 2006, MNRAS, 
371, 983

\bibitem[\protect\citeauthoryear{Suyu et al.}{2014}]{2014ApJ...788L..35S} 
Suyu S.~H., et al., 2014, ApJ, 788, L35 

\bibitem[\protect\citeauthoryear{Suyu et al.}{2013}]{2013ApJ...766...70S} 
Suyu S.~H., et al., 2013, ApJ, 766, 70 

\bibitem[\protect\citeauthoryear{Tagore 
\& Jackson}{2015}]{2015arXiv150500198T} Tagore A.~S., Jackson N., 2015, arXiv, arXiv:1505.00198

\bibitem[\protect\citeauthoryear{Treu}{2010}]{treu} Treu T., 2010, ARA\&A, 48, 87 

\bibitem[\protect\citeauthoryear{Umst{\"a}tter et 
al.}{2005}]{renate} Umst{\"a}tter R., Christensen N., Hendry 
M., Meyer R., Simha V., Veitch J., Vigeland S., Woan G., 2005, PhRvD, 72, 
022001 

\bibitem[\protect\citeauthoryear{Vegetti 
\& Koopmans}{2009}]{2009MNRAS.400.1583V} Vegetti S., Koopmans L.~V.~E., 2009, MNRAS, 400, 1583

\bibitem[\protect\citeauthoryear{Vegetti et 
al.}{2012}]{vegetti1} Vegetti S., Lagattuta D.~J., McKean J.~P., 
Auger M.~W., Fassnacht C.~D., Koopmans L.~V.~E., 2012, Natur, 481, 341 

\bibitem[\protect\citeauthoryear{Vegetti, Czoske, 
\& Koopmans}{2010}]{vegetti3} Vegetti S., Czoske O., Koopmans L.~V.~E., 2010, MNRAS, 407, 225

\bibitem[\protect\citeauthoryear{Vegetti et 
al.}{2010}]{vegetti2} Vegetti S., Koopmans L.~V.~E., Bolton A., 
Treu T., Gavazzi R., 2010, MNRAS, 408, 1969 

\bibitem[\protect\citeauthoryear{Vegetti et 
al.}{2014}]{2014MNRAS.442.2017V} Vegetti S., Koopmans L.~V.~E., Auger 
M.~W., Treu T., Bolton A.~S., 2014, MNRAS, 442, 2017 

\bibitem[\protect\citeauthoryear{Warren 
\& Dye}{2003}]{2003ApJ...590..673W} Warren S.~J., Dye S., 2003, ApJ, 590, 673
\end{thebibliography}
\end{document}